\documentclass[acmsmall]{acmart}

\usepackage{algorithm}
\usepackage{algorithmicx}
\usepackage{algpseudocode}
\usepackage{amsmath}
\usepackage{enumitem}
\usepackage{graphicx}
\usepackage{multirow}
\usepackage{natbib}
\usepackage{tabularx}
\usepackage{booktabs}
\usepackage{setspace}
\usepackage{subcaption}
\usepackage{xspace}
\usepackage{listings}
\usepackage{tcolorbox}
\tcbuselibrary{skins}
\usepackage{balance}
\usepackage[all]{nowidow}

\algrenewcommand\algorithmicrequire{\textbf{Input:} }
\algrenewcommand\algorithmicensure{\textbf{Output:} }
\algnewcommand{\algorithmicand}{\textbf{ and }}
\algnewcommand{\algorithmicor}{\textbf{ or }}
\algnewcommand\AND{\algorithmicand}
\algnewcommand\OR{\algorithmicor}
\newcommand{\FRAM}{FuzzSDN\xspace}
\newcolumntype{Y}{>{\centering\arraybackslash}X}

\title{Learning Failure-Inducing Models for Testing Software-Defined Networks}

\author{Rapha{\"e}l Ollando}
\email{raphael.ollando@uni.lu}
\affiliation{%
	\institution{University of Luxembourg}
	\streetaddress{29 Avenue John F. Kennedy}
	\city{Luxembourg}
	\postcode{1859}
	\country{Luxembourg}
}

\author{Seung Yeob Shin}
\email{seungyeob.shin@uni.lu}
\affiliation{%
	\institution{University of Luxembourg}
	\streetaddress{29 Avenue John F. Kennedy}
	\city{Luxembourg}
	\postcode{1859}
	\country{Luxembourg}
}

\author{Lionel C. Briand}
\email{lionel.briand@lero.ie}
\affiliation{%
	\institution{Lero centre, University of Limerick}
	\streetaddress{Tierney building}
	\city{Limerick}
	\postcode{V94 NYD3}
	\country{Ireland}
}
\affiliation{%
	\institution{University of Ottawa}
	\streetaddress{800 King Edward Avenue}
	\city{Ottawa}
	\postcode{ON K1N 6N5}
	\country{Canada}
}

\newcounter{commentnumber}

\begin{document}

\begin{abstract}
Software-defined networks (SDN) enable flexible and effective communication systems that are managed by centralized software controllers. However, such a controller can undermine the underlying communication network of an SDN-based system and thus must be carefully tested. When an SDN-based system fails, in order to address such a failure, engineers need to precisely understand the conditions under which it occurs.
% <<<< Revision 1
In this article, we introduce a machine learning-guided fuzzing method, named FuzzSDN, aiming at both (1)~generating effective test data leading to failures in SDN-based systems and (2)~learning accurate failure-inducing models that characterize conditions under which such system fails.
To our knowledge, no existing work simultaneously addresses these two objectives for SDNs.
% ===== Revision 1
We evaluate \FRAM by applying it to systems controlled by two open-source SDN controllers. Further, we compare \FRAM with two state-of-the-art methods for fuzzing SDNs and two baselines for learning failure-inducing models. Our results show that (1)~compared to the state-of-the-art methods, \FRAM generates at least 12 times more failures, within the same time budget, with a controller that is fairly robust to fuzzing and (2)~our failure-inducing models have, on average, a precision of 98\% and a recall of 86\%, significantly outperforming the baselines.
\end{abstract}




% It is important to put this just before the \maketitle
% for the TOSEM template
\onecolumn

% Revision 1
% \input{revision_1/letter}
% \input{revision_1/metareview}
% \input{revision_1/reviewer_1}
% \input{revision_1/reviewer_2}
% \input{revision_1/reviewer_3}

% Revision 2
% \input{revision_2/letter}
% \input{revision_2/metareview}

\maketitle

\setcounter{page}{1}

% !TEX root =  ../paper.tex
\section{Introduction}
\label{sec:introduction}

Software-defined networks (SDN)~\cite{SDN:15} have emerged to enable programmable networks that allow system operators to manage their systems in a flexible and efficient way. SDNs have been widely deployed in many application domains, such as data centers~\cite{Drutskoy2013, Wang2017}, the Internet of Things~\cite{Rafique2020,ShinNSB0Z20}, and satellite communications~\cite{Ferrus2016,LiuSZCSK18}. The main idea behind SDNs is to transfer the control of networks from localized, fixed-behavior controllers distributed over a set of network switches (in traditional networks) to a logically centralized and programmable software controller. With complex software being an integral part of SDNs, developing SDN-based systems (SDN-systems), e.g., data centers, entails interdisciplinary considerations, including software engineering.
 
In the context of developing SDN-systems, software testing becomes even more important and challenging when compared to what is required in traditional networks that provide static and predictable operations. In particular, even though the centralized controller in an SDN-system enables flexible and efficient services, it can undermine the entire communication network it manages. A software controller presents new attack surfaces that allow malicious users to manipulate the systems~\cite{Dhawan2015:SPHINX, Ropke2015:Rootkits}. For example, if malicious users intercept and poison communication in the system (using ARP spoofing~\cite{ContiDL16}), such attacks broadly impact the system due to its centralized control. Furthermore, the centralized controller interacts with diverse kinds of components such as applications and network switches, which are typically developed by different vendors. Hence, the controller is prone to receiving unexpected inputs provided by applications, switches, or malicious users, which may cause system failures, e.g., communication breakdown. 

To test an SDN controller, engineers need first to explore its possible input space, which is very large. A controller takes as input a stream of control messages which are encoded according to an SDN communication protocol (e.g., OpenFlow~\cite{OpenFlowSpec}). For example, if a control message is encoded with OpenFlow, it can have $2^{2040}$ distinct values~\cite{OpenFlowSpec}. Second, engineers need to understand the characteristics of test data, i.e., control messages, that cause system failures. However, manually inspecting test data that cause failures is time-consuming and error-prone. Furthermore, misunderstanding such causes typically leads to unreliable fixes. 

% <<<<< Revision 1
There are a number of prior research strands that aim at testing SDN-systems~\cite{NandaZDWY16, BhuniaG17, Jero2017:BEADS, Lee2017:DELTA, Zhang17, WooLKS18, Alshanqiti19, ChicaIB20, AlbabDHKSWTGY22}. Most of them come from the network research field and focus on security testing relying on domain knowledge, e.g., known attack scenarios~\cite{Lee2017:DELTA}.
The most pertinent research works applied fuzzing techniques to different components of SDN-systems.
For example, RE-CHECKER~\cite{WooLKS18} fuzzes RESTful services provided by SDN controllers.
SwitchV~\cite{AlbabDHKSWTGY22} relies on fuzzing and symbolic execution to test SDN switches.
BEADS~\cite{Jero2017:BEADS} tests SDN controllers by being aware of the OpenFlow specification.
However, none of these fuzzing techniques employ interpretable machine learning techniques to guide their fuzzing process and to provide models that characterize failure-inducing conditions. 
Even though the software engineering community has introduced numerous testing methods, testing SDN-systems has gained little attention.
The most pertinent research lines have proposed techniques for learning-based fuzzing~\cite{GodefroidPS17,Chen19,ZhaoLWSH19} and abstracting failure-inducing inputs~\cite{Gopinath2020,KampmannHSZ20} to efficiently explore the input space and characterize effective test data that cause system failures.
Learn@Fuzz~\cite{GodefroidPS17} employs neural-network-based learning methods for building a model of PDF objects for grammar-based fuzzing.
A prior learning-guided fuzzing technique, named smart fuzzing~\cite{Chen19}, relies on deep learning techniques to test systems controlled by programmable logic controllers (PLC).
SeqFuzzer~\cite{ZhaoLWSH19} uses deep learning techniques to infer communication protocols underlying PLC systems and generate fuzzed messages.
However, deep learning techniques are not suitable to characterize failure-inducing test data as they do not provide interpretable models.
Furthermore, these techniques do not account for the specificities of SDNs, such as SDN architecture and communication protocol.
% ===== Revision 1
Existing work on abstracting failure-inducing inputs~\cite{Gopinath2020,KampmannHSZ20} targets software programs that take as input strings such as command-line utilities (e.g., find and grep), which are significantly different from SDN-systems.
In summary, no existing work simultaneously tackles the problem of efficiently exploring the input space and accurately characterizing failure-inducing test data while accounting for the specificities of SDNs.

\textbf{Contributions.} In this article, we propose \FRAM, a machine learning-guided \underline{Fuzz}ing method for testing \underline{SDN}-systems. In particular, \FRAM targets software controllers deployed in SDN-systems. \FRAM relies on fuzzing guided by machine learning (ML) to both (1)~efficiently explore the test input space of an SDN-system's controller (generate test data leading to system failures) and (2)~learn failure-inducing models that characterize input conditions under which the system fails. This is done in a synergistic manner where models guide test generation and the latter also aims at improving the models. A failure-inducing model is practically useful~\cite{Gopinath2020} for the following reasons: (1)~It facilitates the diagnosis of system failures. \FRAM provides engineers with an interpretable model specifying how likely are failures to occur, e.g., the system fails when a control message is encoded using OpenFlow V1.0 and contains IP packets, thus providing concrete conditions under which a system will probably fail. Such conditions are much easier to analyze than a large set of individual failures. (2)~A failure-inducing model enables engineers to validate their fixes. Engineers can fix and test their code against the generated test data set. A failure-inducing model can also be used as a test data generator to reproduce the system failures captured in the model. Hence, engineers can \hbox{better validate their fixes using an extended test data set.}

We evaluated \FRAM by applying it to several systems controlled by well-known open-source SDN controllers: ONOS~\cite{Berde2014:ONOS} and RYU~\cite{RYU}. In addition, we compared \FRAM with two state-of-the-art methods (i.e., DELTA~\cite{Lee2017:DELTA} and BEADS~\cite{Jero2017:BEADS}) that generate test data for SDN contollers and two baselines that learn failure-inducing models. As baselines, we extended DELTA and BEADS to produce failure-inducing models, since they were not originally designed for that purpose but were nevertheless our best options. Our experiment results show that, compared to state-of-the-art methods, FuzzSDN generates at least 12 times more failing control messages, within the same time budget, with a controller that is fairly robust to fuzzing. \FRAM also produces accurate failure-inducing models with, on average, a precision of 98\% and a recall of 86\%, which significantly outperform models inferred by the two baselines. Furthermore, \FRAM produces failure-inducing conditions that are consistent with those reported in the literature~\cite{Jero2017:BEADS}, indicating that \FRAM is a promising solution for automatically characterizing failure-inducing conditions and thus reducing the effort needed for manually analyzing test results. Last, \FRAM is applicable to systems with large networks as its performance does not depend on network size. Our detailed evaluation results and the \hbox{\FRAM tool are available online~\cite{Artifacts}.}

\noindent\textbf{Organization.} The rest of this article is organized as follows: Section~\ref{sec:problem} introduces the background and defines the specific problem of learning failure-inducing models for testing SDN-systems. Section~\ref{sec:approach} describes \FRAM. Section~\ref{sec:evaluation} evaluates \FRAM in a large empirical study. Section~\ref{sec:related work} compares \FRAM with related work. Section~\ref{sec:conclusion} concludes this article.
% !TEX root =  ../paper.tex
\section{Background and Problem Description}
\label{sec:problem}

In this section, we describe the fundamental concepts of an SDN-system. We then discuss the problem of identifying input conditions under which the system under test fails.

% <<<<< Revision 2
\begin{figure*}[t]
	\centering
	\includegraphics[width=0.5\textwidth]{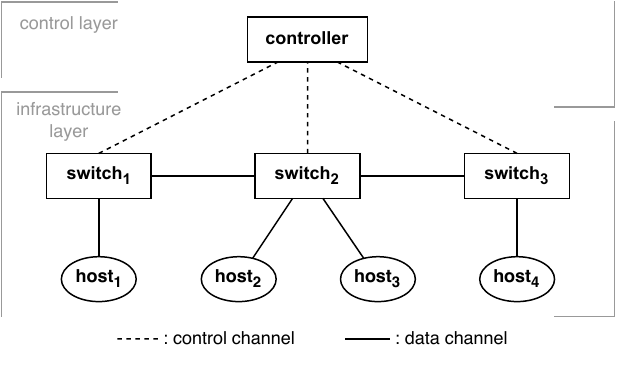}
	\caption{An SDN topology example containing one controller, three switches, and four hosts.}
	\label{fig: SDN System topology}
\end{figure*}
% ===== Revision 2

\noindent\textbf{Controller.} In general, an SDN-system is composed of two layers: infrastructure and control layers. The infrastructure layer contains physical components, such as switches and links, that build a physical network transferring data flows such as audio and video data streams. The control layer is a software component developed by software engineers to meet the system's requirements. Specifically, the software controller manages the physical components in the infrastructure layer to implement, for example, system-specific data forwarding, security management, and failure recovery algorithms.
% <<<<< Revision 2
Figure~\ref{fig: SDN System topology} shows an SDN topology example, containing a controller in the control layer along with three switches and four hosts (such as servers and clients) in the infrastructure layer.
Note that this controller is the target of our testing, which is not present in traditional networking systems.
% ===== Revision 2
In this article, we simplified the SDN layered architecture~\cite{SDN:15} to clearly explain our contributions by abstracting out network-specific details, e.g., SDN southbound interfaces, that are not important to present this work.

% <<<<< Revision 1
In traditional networking systems, controllers are typically embedded in the physical devices, operating based on their local configurations.
This decentralized control approach contrasts with the centralized control of SDN-systems.
The SDN architecture centralizes decision-making, aiming for enhanced efficiency and dynamic network management.
However, this can also lead to SDN-specific failures caused by such central point of control.
% ===== Revision 1

\noindent\textbf{Message.} The SDN controller and switches in a network communicate by exchanging control messages. A control message is encoded as a sequence of values following a specific communication protocol. For instance, OpenFlow~\cite{OpenFlowSpec} is a de facto standard communications protocol used in many SDN-systems~\cite{Lara2014}, enabling communication between the control and infrastructure layers. To exercise the behavior of the controller under test, therefore, testing explores the space of possible control messages.

% <<<<< Revision 1
In traditional networking systems, communication between devices usually employs well-established transport protocols, such as UDP~\cite{UDPSpec} and TCP~\cite{TCPSpec}, for data transmission.
For exchanging routing and state information, routers and switches use standard routing protocols, such as OSPF~\cite{OSPFSpec} and BGP~\cite{BGPSpec}.
Since these protocols were designed to serve specific purposes in the context of traditional static networks, the messages they encode are more similar and simpler than those used in SDN.
In contrast, SDN messages, such as those in OpenFlow, carry a variety of data including flow setups, modifications, and statistics, allowing for more flexibility in network services.
However, this also entails a wider range of potential failures that should be accounted for during testing.
% ===== Revision 1

\noindent\textbf{Failure.}
% <<<<< Revision 1
Like other software components, SDN controllers may have faults that can lead to service failures perceivable by users.
These failures can manifest in various forms in the context of SDN-systems.
Specifically, previous studies on SDN testing~\cite{Lee2017:DELTA,Jero2017:BEADS,Alshanqiti19,LeeWKYPS20,ShuklaSSCZF20} have investigated the following failures relevant to SDN controllers:
(1)~Controller-switch disconnection. When communications between the controller and the switches are unexpectedly disconnected, the system obviously cannot operate as intended.
For example, if a switch does not receive timely commands from the controller due to this disconnection, the switch relies on outdated forwarding rules installed earlier, possibly leading to dropping data flows entirely.
(2)~SDN operation stall. This failure refers to situations where the required execution of an SDN operation is unexpectedly prevented or delayed.
For example, an SDN controller might stall the installation of forwarding rules, affecting the construction of the communication path in the SDN.
(3)~Incorrect understanding of network status. SDN controllers rely on a centralized view of the network to make control decisions.
If this centralized view becomes inconsistent with the actual network, regardless of the reasons, e.g., receiving outdated or incorrect data from the network switches, the controller may fail to make decisions that reflect the network's actual conditions.
For example, if a link between two switches is broken, but the controller's view still considers it operational, the controller fails to instruct the switches to reroute data flows around the broken link, resulting in dropped data flows.
(4)~Overutilization of resources. SDN controllers may overutilize their processing and memory units due to various reasons, such as handling an unexpectedly high volume of requests.
For example, such a failure caused by a distributed denial-of-service (DDoS) attack~\cite{MirkovicR2004} targeting the SDN controller can lead to slower response times and potential service outages.
We note that, in addition to failures specific to SDN controllers, failures observed in traditional networks, such as communication breakdowns among hosts (e.g., data servers and clients) and network performance degradation, are also relevant in SDN-systems as its infrastructure layer usually employs traditional network protocols, such as TCP, UDP, and IP.
% ===== Revision 1

% <<<<< Revision 1
In traditional networking systems, failures tend to be localized.
If a router or switch fails, only its directly connected neighbours (i.e., localized segments of the network) are affected.
Similarly, if a server or client fails, only the applications and users directly reliant on that specific device are impacted.
In contrast, SDN-systems have a centralized failure point in the controller.
If the controller crashes or loses connection with the switches, the entire network can be affected.
This central point of potential failure makes rigorous testing essential to ensure robustness and reliability.
% ===== Revision 1

\noindent\textbf{Problem.} When developing and operating an SDN-system, engineers must handle system failures that are triggered by unexpected control messages. In particular, engineers need to ensure that the system behaves in an acceptable way in the presence of failures. In an SDN-system, its controller is prone to receiving unexpected control messages from switches in the system~\cite{ChicaIB20}. For example, network switches, which are typically developed by different vendors, may send control messages that fall outside the scope of the controller's expectations. Such unexpected messages can also be sent by the switches due to various reasons such as malfunctions and bugs in the switches, as well as inconsistent implementations of a communication protocol between the controller and switches~\cite{LeeWKYPS20}. Furthermore, prior security assessments of SDNs have found several attack surfaces, leading to vulnerable applications and communications protocols that enable malicious actors to send manipulated messages over an SDN~\cite{Lee2017:DELTA,Jero2017:BEADS}.

When a failure occurs in an SDN-system, engineers need to determine the conditions under which such failures occur. These conditions define a set of control messages that cause the failure. Identifying such conditions in a precise and interpretable form is in practice useful, as it enables engineers to diagnose the failure with a clear understanding of the conditions that induce it. In addition, engineers can produce an extended set of control messages by utilizing the identified conditions to test the system after making changes to address the failure. In general, any fix should properly address other control messages that induce the same failure. Our work aims to both effectively test an SDN-system's controller by identifying control messages that lead to system failures, and then automatically identify an accurate failure-inducing model that characterizes conditions under which the SDN-system fails. Such conditions define a set of failure-inducing control messages.

% !TEX root =  ../paper.tex
\section{Approach}
\label{sec:approach}

\begin{figure}[t]
	\centering
	\includegraphics[width=0.7\columnwidth]{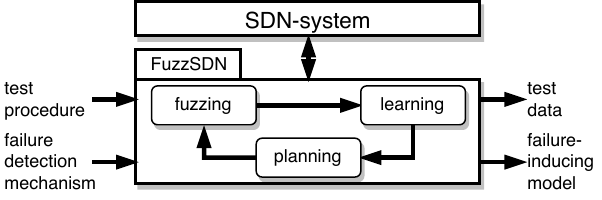}
	%\vspace{-0.4em}
	\caption{An overview of our ML-guided \underline{Fuzz}ing method for testing \underline{SDN}-systems (\FRAM).}
	\label{fig:approach_overview}
%\vspace{-1.0em}
\end{figure}

Figure~\ref{fig:approach_overview} shows an overview of our ML-guided \underline{Fuzz}ing method for testing \underline{SDN}-systems (\FRAM).
Specifically, \FRAM relies on fuzzing and ML techniques to effectively test an SDN-system's controller and generate a failure-inducing model that characterizes conditions under which the system fails. As shown in Figure~\ref{fig:approach_overview}, \FRAM takes as input a test procedure and a failure detection mechanism defined by engineers.

% <<<<< Revision 1
A test procedure consists of three steps: initializing the SDN-system, executing the test scenario, and tearing down the system.
In the initialization step, the procedure configures the entities in the SDN-system, such as the controller, switches, and connections, bringing the system to a desired (ready) state for executing the test scenario.
For example, during the initialization, the procedure might configure switches with empty or preinstalled forwarding tables, which define how to forward data packets through the SDN. This system configuration allows for testing how the controller instructs the switches in the given setting.
The test scenario represents a specific operation (use case) of the SDN-system, involving the exchange of control messages between the controller and switches that \FRAM can fuzz.
For example, if the test scenario specifies a data transmission from one host to another connected to the SDN, executing the test scenario leads to the exchange of control messages between the controller and switches.
These messages are exchanged to discover the locations of the two hosts in the SDN and to instruct the relevant switches with the proper forwarding tables to enable data transmission.
In the teardown step, the procedure resets the SDN-system to its default (pristine) state after executing the test scenario, since \FRAM requires multiple independent executions of the test procedure.
% ===== Revision 1
% Based on the input test procedure, we note that the system generates control messages that \FRAM fuzzes.

% <<<<< Revision 1
A failure detection mechanism acts as a test oracle, determining whether the system fails or successfully completes the given test procedure.
For example, depending on the test procedure, it detects instances of communication breakdown, controller crashes, or performance degradation.
Since an SDN-system provides monitoring tools that enable engineers to oversee system behavior and performance, implementing such a failure detection mechanism is straightforward.
% ===== Revision 1

\FRAM then outputs test data and a failure-inducing model. The data includes the set of control messages that induced the failure detected by the failure detection mechanism. A failure-inducing model abstracts such test data in the form of conditions and probabilities pertaining to the failure.

As shown in Figure~\ref{fig:approach_overview}, \FRAM realizes an iterative process consisting of the following three steps: (1)~The fuzzing step sniffs and modifies a control message passing through the control channel from the SDN switches to the controller. The fuzzing step repeats the execution of the input test procedure and modifies only one selected control message for each execution of the system. It then produces a labeled dataset that associates fuzzing outputs (i.e., modified control messages) and their consequences in the system (i.e., system failure or success).
%Note that fuzzing multiple control messages at a time poses new challenges due to, e.g., message dependencies and internal state changes in the system. Addressing such challenges is beyond the scope of this paper.
(2)~The learning step takes as input the labeled dataset created by the fuzzing step and uses a supervised learning technique, e.g., RIPPER~\cite{Cohen1995:FERI}, to create a failure-inducing (classification) model. This model identifies a set of control messages that cause the failure defined in the failure detection mechanism. Over the iterations of \FRAM, such models are used to guide the behavior of the fuzzing step in the next iteration.
(3)~The planning step instructs the fuzzing step based on the failure-inducing model created by the learning step. Following such instructions, the fuzzing step then efficiently explores the space of control messages to be fuzzed and adds new data points (i.e., modified control messages and their consequences in the system) to the existing labeled dataset. The updated dataset is then used by the learning step to produce an improved failure-inducing model. \FRAM stops the iterations of the fuzzing, learning, and planning steps when the accuracy of the output failure-inducing model reaches an acceptable level or the execution time of \FRAM exceeds an allotted time budget. Below, we explain each step of \FRAM in detail.

% <<<<< Revision 1
We assume that engineers using \FRAM have expertise in SDN, thus equipping them with the capability to provide required inputs: test procedures and failure detection mechanisms.
They should also possess knowledge of the SDN protocol to understand a failure-inducing model that characterizes a set of failure-inducing control messages.
Such assumptions are reasonable since engineers need to devise test procedures and failure detection mechanisms when testing their SDN-systems, independently of \FRAM, and must know the SDN protocol as it defines the input space of the SDN controller under test.
% ===== Revision 1

\subsection{Fuzzing step: Initial fuzzing}
\label{subsec:fuzzing step:initial}

During the fuzzing step, \FRAM manipulates control messages in an SDN-system to cause a system failure. To do so, \FRAM utilizes a man-in-the-middle attack, which is a well-known security attack technique in the network domain~\cite{ContiDL16}. The attack technique enables \FRAM to intercept control messages transmitting through the control channel and inject modified messages. In addition, \FRAM pretends to be both legitimate participants (i.e., SDN switches and controllers) of the control channel. Hence, the system under test is not aware of \FRAM while it is running. We omit network-specific details of the attack technique, as they are not part of our contributions; instead, we refer interested readers to the relevant literature~\cite{ContiDL16}.

\begin{figure*}[t]
	\centering
        \includegraphics[width=\linewidth]{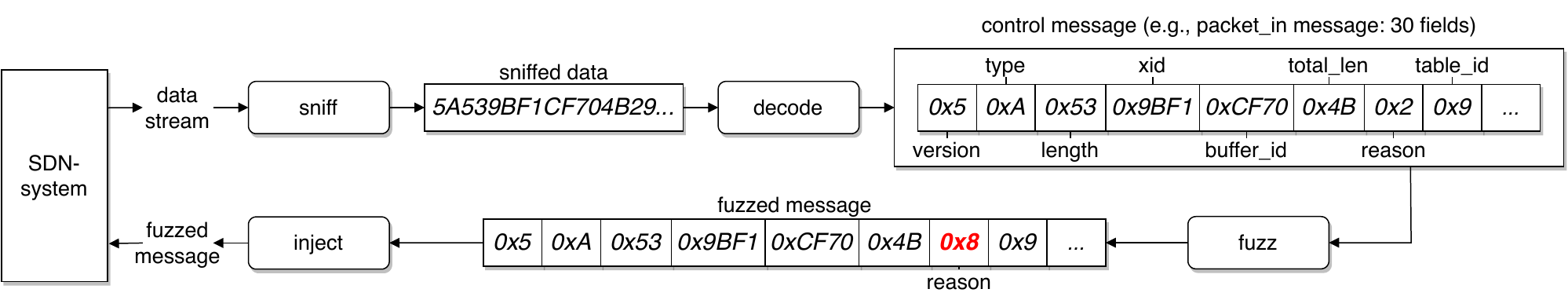}
%\vspace{-1.5em}
	% \caption{A data flow example of fuzzing a control message.}
        \caption{A data flow example of fuzzing a control message (e.g., packet\_in message).}
	\label{fig:fuzzing flow}
%\vspace{-0.5em}
\end{figure*}

Figure~\ref{fig:fuzzing flow} shows a data flow example that illustrates how our fuzzing technique manipulates a control message. As the control channel in the SDN-system transmits a data stream, the fuzzing step first sniffs it as a byte string. It then decodes the sniffed string according to the adopted SDN protocol (e.g., OpenFlow~\cite{OpenFlowSpec}) to identify a control message to be fuzzed.
In Figure~\ref{fig:fuzzing flow}, the byte string 0x5A539BF1CF704B29 is sniffed and is then identified as a packet\_in control message encoded in 30 fields according to OpenFlow. Note that a packet\_in message is one of the control messages sent by a switch to a controller in order to notify that the switch receives a packet. For details of the packet\_in message, we refer readers to the OpenFlow specification~\cite{OpenFlowSpec}.

\FRAM fuzzes the control message by accounting for the syntax requirements (i.e., grammar) defined in the SDN protocol and then injects the fuzzed message into the control channel in the system. We note that \FRAM could apply a simple random fuzzing method that replaces the sniffed string with a random string. However, the majority of byte strings generated by random fuzzing would be invalid control messages that would be immediately rejected by the SDN message parser in the system~\cite{Jero2017:BEADS}. \FRAM therefore accounts for the SDN protocol in order to generate valid control messages that test software components beyond the message parsing layer of the system, which is a desirable feature in practice~\cite{Jero2017:BEADS}.

%\begingroup
%\setlength{\textfloatsep}{0pt}
%\setlength{\intextsep}{0pt}
\begin{algorithm}[t]
	\caption{Initial fuzzing}
	\label{alg:initial fuzzing}
	%\footnotesize
	\begin{algorithmic}[1]
		\Require
		\Statex $\mathit{msg}$: control message to be fuzzed
		\Ensure
		\Statex $\mathit{msg^\prime}$: control message after fuzzing
		\Statex
		% \State $F \gets \mathit{select{\mathunderscore}fields}(\mathit{msg})$
            \State $F \gets \mathit{select{\mathunderscore}rand{\mathunderscore}fields}(\mathit{msg})$
		\State $\mathit{msg}^\prime \gets \mathit{msg}$
		\ForAll{$f \in F$}
		% \State $\mathit{msg}^\prime \gets \mathit{replace}(\mathit{msg^\prime}, f, \mathit{rand}(f))$
            \State $\mathit{msg}^\prime \gets \mathit{replace}(\mathit{msg^\prime}, f, \mathit{rand{\mathunderscore}valid}(f))$
		\EndFor
		\State \Return $\mathit{msg^\prime}$
	\end{algorithmic}
\end{algorithm}
%\endgroup

\noindent\textbf{Initial fuzzing.} At the first iteration of \FRAM, since a failure-inducing model is not present, the fuzzing step behaves as described in Algorithm~\ref{alg:initial fuzzing}. Given a control message $\mathit{msg}$, the algorithm modifies it and returns a fuzzed message $\mathit{msg^\prime}$. As shown on line 1, the algorithm first randomly selects a set $F$ of fields in $\mathit{msg}$. For each field $f$ in $F$, the algorithm replaces the original value of $f$ with a new value randomly selected from its value range (lines 2-5).
% <<<<< Revision 2
Hence, Algorithm~\ref{alg:initial fuzzing} operates in linear time, relative to the number of message fields ($|F|$).
% ===== Revision 2
Note that our ML-guided fuzzing method is described in Section~\ref{subsec:fuzzing:guided}.
For example, in Figure~\ref{fig:fuzzing flow}, the algorithm modifies the sniffed packet\_in message by replacing the reason field value 0x2 with 0x8, which is randomly chosen within its value range.

\noindent\textbf{Data collection.} To generate a failure-inducing model, \FRAM uses a supervised ML technique~\cite{WittenFH16} that requires a labeled dataset the fuzzing step generates. Specifically, at each iteration of \FRAM, the fuzzing step is executed $n$ times based on an allotted time budget. Each execution of the fuzzing step (re)runs the input test procedure (Figure~\ref{fig:approach_overview}) and modifies a control message. \FRAM then monitors the system response to the modified control message $\mathit{msg}$ using the failure detection mechanism (Figure~\ref{fig:approach_overview}). We denote by $\mathit{presence}$ (resp. $\mathit{absence}$) the label indicating that the failure is present (resp. absent) in the system response. For each iteration $i$ of \FRAM, the fuzzing step creates a labeled dataset $D_i$ by adding $n$ tuples $(\mathit{msg}_1,l_1)$, $\ldots$, $(\mathit{msg}_n,l_n)$ to $D_{i-1}$, where a label $l_j$ could be either $\mathit{presence}$ or $\mathit{absence}$ and $D_0$ $=$ $\{\}$. We note that \FRAM uses an accumulated dataset $D$ $=$ $D_1$ $\cup$ $\ldots$ $\cup$ $D_i$ to infer a failure-inducing model \hbox{at each iteration $i$.}

\subsection{Learning step}
\label{subsec:learning step}

We cast the problem of learning failure-inducing models into a binary classification problem in ML. Given a labeled dataset obtained from the fuzzing step, the learning step infers a prediction model, i.e., a failure-inducing model, that classifies a control message as either the $\mathit{presence}$ or $\mathit{absence}$ class. A classification result predicts whether or not a control message induces a system failure as captured by the detection mechanism.

\FRAM aims at providing engineers not only with failure-inducing control messages but also accurate conditions under which the system fails (as described in Section~\ref{sec:introduction}). Hence, we opt to use a learning technique that produces an interpretable model~\cite{Molnar2022}. Engineers could use the prediction results to identify a set of control messages predicted to induce system failures as a test suite for testing the system. An interpretable model would help engineers figure out why the failure occurs in the system. We encode fields (e.g., version, type, and length in Figure~\ref{fig:fuzzing flow}) of a control message as features in a labeled dataset so that an interpretable ML technique builds a failure-inducing model using fields as features.
For example, such a model could explain that the system fails when receiving a control message with an incompatible version number, which is a higher version than the system supports.
%\textcolor{orange}{For example, a model which contains a rule "$\mathit{version}$ $\geq$ $7$ => class=presence" indicates that the system fails when receiving a control message with version number above 6 (0x06), which is a version that do not exists in the OpenFlow Protocol}

\FRAM employs RIPPER (Repeated Incremental Pruning to Produce Error Reduction)~\cite{Cohen1995:FERI}, an interpretable rule-based classification algorithm, to learn a failure-inducing model. We opt to use RIPPER as it generates pruned decision rules that are more concise and thus interpretable than commonly used decision trees (e.g., C4.5~\cite{Quinlan1993}), which are prone to the replicated subtree problem~\cite{WittenFH16}. Further, RIPPER has been successfully applied to many software engineering problems involving classification and rule inference~\cite{HaqSNB21,GhotraMH15,Brindescu0LS20}. Given a labeled dataset $D$, RIPPER produces a set $R$ of decision rules. A decision rule is a simple IF-condition-THEN-prediction statement, consisting of a condition on the fields of a control message (e.g., $\mathit{version}$ $>$ $5$ $\wedge$ $\mathit{length}$ $\geq$ $10$) and a prediction indicating either the $\mathit{presence}$ or $\mathit{absence}$ class.

For a decision rule, the learning step measures a confidence score~\cite{WittenFH16} to estimate the accuracy of the rule in predicting the actual class of control messages that satisfy the rule's condition. Specifically, given a labeled dataset $D$ and a decision rule $r$, we denote by $t$ the total number of control messages in $D$ that satisfy the condition of $r$, i.e., the condition is evaluated to true. Among these $t$ control messages, often, there are some control messages whose labels defined in $D$ do not match $r$'s prediction. We denote by $f$ the number of such control messages. The learning step computes a confidence score $c(r)$ of $r$ by $c(r)$ $=$ $(t-f)/t$.
For example, given a rule $r$: IF $\mathit{version} > 5 \wedge \mathit{length} \geq 10$ THEN $\mathit{class} = \mathit{presence}$, suppose 88 control messages in a labeled dataset $D$ satisfy the condition of $r$ and 7 out of the 88 control messages are labeled with $\mathit{absence}$. Then, the confidence score $c(r)$ is $(88-7)/88 = 0.92$.
\FRAM uses the confidence score $c(r)$ in the planning \hbox{step to guide our fuzzing strategy.}

\subsection{Planning step}
\label{subsec:planning step}

The planning step guides fuzzing based on a failure-inducing model, i.e., a set $R$ of decision rules inferred from the learning step. Given an original control message to be fuzzed, the main idea is to generate a set of modified control messages using $R$. To this end, the planning step exploits these decision rules to generate effective control messages that induce failures.

%\begingroup
%\setlength{\textfloatsep}{0pt}
%\setlength{\intextsep}{0pt}
\begin{algorithm}[t]
	\caption{Planning}
	\label{alg:planning}
	%\footnotesize
	\begin{algorithmic}[1]\onehalfspacing
		\Require
		\Statex $D$: labeled dataset
		\Statex $R$: decision rules
		\Statex $n$: number of control messages to be fuzzed
		\Ensure
		\Statex $B$: budget distribution to the decision rules $R$
		\Statex
		\State \textcolor{gray}{//count numbers of majorities and minorities in $D$}
		\State $\mathit{minor} \gets \mathit{get\mathunderscore num\mathunderscore minorities}(D)$
		\State $\mathit{major} \gets \mathit{get\mathunderscore num\mathunderscore majorities}(D)$
		\State \textcolor{gray}{//estimate numbers of majorities and minorities after the next iteration}
		\State $\mathit{minor}^\prime \gets \min((|D|+n)/2 - \mathit{minor}, n)$
		\State $\mathit{major}^\prime \gets n - \mathit{minor}^\prime$
		\State \textcolor{gray}{//assign budgets to minority rules}
		\State $B \gets \{\}$
		\State $R^{\mathit{minor}} \gets \mathit{get\mathunderscore minority\mathunderscore rules}(R)$
		\ForAll{$r \in R^{\mathit{minor}}$}
		\State $b \gets \mathit{minor}^\prime \times c(r) / \mathit{sum\mathunderscore c}(R^{\mathit{minor}})$
		\State $B \gets B \cup (r,b)$
		\EndFor
		\State \textcolor{gray}{//assign budgets to majority rules}
		\State $R^{\mathit{major}} \gets \mathit{get\mathunderscore majority\mathunderscore rules}(R)$
		\ForAll{$r \in R^{\mathit{major}}$}
		\State $b \gets \mathit{major}^\prime \times c(r) / \mathit{sum\mathunderscore c}(R^{\mathit{major}})$
		\State $B \gets B \cup (r,b)$
		\EndFor
		\State \Return $B$
	\end{algorithmic}
\end{algorithm}
%\endgroup

\noindent\textbf{Imbalance handling.} Algorithm~\ref{alg:planning} describes how the planning step uses the set $R$ of decision rules inferred in the current iteration of \FRAM to guide the fuzzing step for the next iteration. We note that the planning step accounts for the imbalance problem~\cite{WittenFH16} that usually causes poor performance of ML algorithms. In a labeled dataset, when the number of data instances of one class is much higher than such number for another class, ML classification models have a tendency to predict the majority class. In our study, such models are not practically useful, as engineers are more interested in control messages that cause system failures.

As shown on lines 1-3 of Algorithm~\ref{alg:planning}, the planning step first counts the number $\mathit{minor}$ (resp. $\mathit{major}$) of minority (resp. majority) control messages in the given labeled dataset $D$. Given the number $n$ of control messages to be fuzzed in the next iteration, lines 4-6 of the algorithm then estimate the number $\mathit{minor}^\prime$ (resp. $\mathit{major}^\prime$) of control messages associated with the minority (resp. majority) class to be added to $D$ to create a balanced dataset, containing $|D|+n$ control messages. Specifically, \FRAM needs $(|D|+n)/2 - \mathit{minor}$ control messages associated with the minority class to balance $D$ in the next iteration.
Table~\ref{tbl:planning_example} shows examples of $\mathit{minor}$, $\mathit{major}$, $\mathit{minor}^\prime$, and $\mathit{major}^\prime$ at each iteration of \FRAM computed by Algorithm~\ref{alg:planning}.
For example, at the first iteration of \FRAM, when $\mathit{minor} = 10$, $\mathit{major} = 190$, and the number of control messages to be fuzzed $n = 200$, $\mathit{minor}^\prime$ is calculated as  $(200+200)/2 - 10$ $=$ $190$ and $\mathit{major}^\prime = 10$.
%Note that if $(|D|+n)/2 - \mathit{minor}$ is greater than the number $n$ of control messages to be fuzzed, the algorithm aims at adding $n$ control messages associated with the minority class to $D$ in the next iteration.

\begin{table}[t]
\caption{Examples of $\mathit{minor}$, $\mathit{major}$, $\mathit{minor}^\prime$, and $\mathit{major}^\prime$ computed by Algorithm~\ref{alg:planning}, when the number $n$ of control messages to be fuzzed is 200.}
\label{tbl:planning_example}
\centering
%\footnotesize
%>{\hsize=.5\hsize}
\begin{tabularx}{\columnwidth}{X Y Y Y Y Y}
\toprule
iteration & $|D|$ & $\mathit{minor}$ & $\mathit{major}$ & $\mathit{minor}^\prime$ & $\mathit{major}^\prime$ \\
\midrule
1   & 200  & 10  & 190 & 190 & 10 \\
2   & 400  & 125 & 275 & 175 & 25 \\
3   & 600  & 248 & 352 & 152 & 48 \\
4   & 800  & 380 & 420 & 120 & 80 \\
5   & 1000 & 495 & 505 & 105 & 95 \\
6   & 1200 & 600 & 600 & 100 & 100 \\
\bottomrule
\end{tabularx}
\end{table}

\noindent\textbf{Budget distribution.} Lines 7-20 of Algorithm~\ref{alg:planning} describe how the planning step distributes the number $n$ of control messages to be fuzzed in the next iteration to each decision rule $r \in R$. As shown on lines 9-13 of the algorithm, for each rule $r \in R^{\mathit{minor}}$ associated with the minority class, the planning step decides to use $r$ for fuzzing based on its relative confidence score and the number $\mathit{minor}^\prime$ of control messages estimated on line 5. Specifically, the planning step associates $r$ with the number of times $r$ will be applied to fuzz control messages, i.e., $\mathit{minor}^\prime \times c(r) / \mathit{sum\mathunderscore c}(R^{\mathit{minor}})$, where $c(r)$ denotes the rule's confidence score (described in Section~\ref{subsec:learning step}) and $\mathit{sum\mathunderscore c}(R^{\mathit{minor}})$ is defined by $\sum_{r \in R^{\mathit{minor}}}c(r)$, the sum of confidence scores of the rules in $R^{\mathit{minor}}$. The algorithm therefore weighs the rules according to their confidence scores in order to maximize the chance of correct predictions. When fuzzing a control message guided by a rule $r$ with a high confidence score (e.g., 0.99), the system response to the fuzzed control message would highly likely match the prediction of $r$. Lines 14-19 describe how the planning step handles rules associated with the majority class, which is the same as on lines 9-13.
For example, at the first iteration of \FRAM shown in Table~\ref{tbl:planning_example}, let $R$ be $\{r_1, r_2, r_3\}$ where $r_1$ and $r_2$ are associated with the minority class (e.g., $\mathit{presence}$) and $r_3$ with the majority class (e.g., $\mathit{absence}$). Given $R$, if $c(r_1) = 0.8$, $c(r_2) = 0.7$, and $c(r_3) = 0.8$, then Algorithm~\ref{alg:planning} distributes $\mathit{minor}^\prime = 190$ (resp. $\mathit{major}^\prime = 10$) to $r_1$ and $r_2$ (resp. $r_3$) as follows: $190 \times 0.8 / (0.8 + 0.7) = 101$ and $190 \times 0.7 / (0.8 + 0.7) = 89$ (resp. $10 \times 0.8 / 0.8 = 10$). For the second iteration, \FRAM then plans to apply $r_1$ 101 times, $r_2$ 89 times, and $r_3$ 10 times to fuzz control messages.
% <<<<< Revision 2
Algorithm~\ref{alg:planning} operates in linear time, relative to the number of rules ($|R|$), inferred by the learning step.
% ===== Revision 2

We note that during early iterations of \FRAM, the obtained datasets are likely imbalanced because control messages causing system failures are typically difficult to discover via purely random fuzzing, since most fuzzed messages are detected and addressed by the system under test to prevent such failures (see Table~\ref{tbl:planning_example} and our experiment results in Section~\ref{subsec:results}). In addition, due to the small sizes of training datasets in early iterations of \FRAM, RIPPER is often not able to produce accurate failure-inducing models. But as \FRAM continuously iterates the three steps within an allotted time budget, according to Algorithm~\ref{alg:planning}, training datasets are becoming more balanced and larger (see Table~\ref{tbl:planning_example}), enabling RIPPER to produce increasingly accurate failure-inducing models. Furthermore, given $S$ the space of all possible control messages, once a dataset is balanced, the algorithm enables the fuzz step to explore not only the space $P$ of control messages that likely cause failures but also the remaining space $S \setminus P$ of control messages. Note that RIPPER infers a set of rules' conditions (which define $P$): $r_1$, $\ldots$, $r_k$, that are associated with the minority class and a single rule's condition (which define $S \setminus P$) in the form of $\neg r_1$ $\wedge$ $\ldots$ $\wedge$ $\neg r_k$ for the majority class.

\noindent\textbf{Progress monitoring.} To monitor the progress of \FRAM, the planning step uses the standard precision and recall metrics~\cite{WittenFH16} (described in Section~\ref{subsec:setup}). In our context, a high level of precision indicates that the inferred failure-inducing model is able to accurately predict the failure of interest. A failure-inducing model with high recall indicates that most of the control messages actually inducing the failure satisfy the failure-inducing conditions in the model. Hence, a failure-inducing model with a high level of precision and recall is desirable. To compute precision and recall values, \FRAM uses the 10-fold cross-validation technique~\cite{WittenFH16}. In 10-fold cross-validation, a dataset $D$ is split into 10 equal-size folds. Nine folds are used as a training dataset and the other one fold is retained as a test dataset. This process is thus repeated 10 times to compute precision and recall values.

\subsection{Fuzzing Step: ML-guided Fuzzing}
\label{subsec:fuzzing:guided}

From subsequent iterations of \FRAM, the fuzzing step utilizes a set $R$ of decision rules inferred by the learning step according to a budget distribution $B$ computed by the planning step. Using a rule $r \in R$, the fuzzing step modifies a sniffed control message to satisfy the condition of $r$. Further, the fuzzing step employs a mutation operator to diversify fuzzed control messages beyond those restricted by $R$. Below we describe the fuzzing step in detail.

%\begingroup
%\setlength{\textfloatsep}{0pt}
%\setlength{\intextsep}{0pt}
\begin{algorithm}[t]
	\caption{ML-guided Fuzzing}
	\label{alg:guided fuzzing}
	%\footnotesize
	\begin{algorithmic}[1]
		\Require
		\Statex $\mathit{msg}$: control message to be fuzzed
		\Statex $B$: budget distribution to the decision rules $R$
		\Statex $\mathit{mu}$: mutation rate
		\Ensure
		\Statex $\mathit{msg}^\prime$: control message after fuzzing
		\Statex $B^\prime$: budget distribution after fuzzing
		\Statex
		\State \textcolor{gray}{//fuzz a control message based on a rule}
		\State $(r,b) \in B$
		\State $F \gets \mathit{get\mathunderscore fields}(r, \mathit{msg})$
		\State $F^\prime \gets \mathit{solve}(r)$
		\State $\mathit{msg}^\prime \gets \mathit{replace}(\mathit{msg}, F, F^\prime)$
		\State \textcolor{gray}{//update the budget distribution}
		\State $B^\prime \gets B \setminus \{(r,b)\}$
		\If{$b-1 > 0$}
		\State $B^\prime \gets B \cup \{(r,b-1)\}$
		\EndIf
		\State \textcolor{gray}{//mutate the fuzzed control message}
		\ForAll{$f \in \mathit{all\mathunderscore fields}(\mathit{msg}) \setminus F$}
		\If{$\mathit{rand}(0,1) \le \mathit{mu}$}
		\State $\mathit{msg}^\prime \gets \mathit{replace}(\mathit{msg}^\prime, f, \mathit{rand}(f))$
		\EndIf
		\EndFor
		\State \Return $\mathit{msg^\prime}$, $B^\prime$
	\end{algorithmic}
\end{algorithm}
%\endgroup

%\textbf{Rule-based fuzzing.}
Algorithm~\ref{alg:guided fuzzing} describes a fuzzing procedure that modifies a sniffed control message $\mathit{msg}$ using a budget distribution $B$. As shown on lines 1-5 of the algorithm, the fuzzing step first chooses a budget assignment $(r,b) \in B$, where $r$ denotes a rule to apply in fuzzing, and $b$ denotes how many times the rule $r$ will be exploited by the fuzzing step. Given the rule $r$, the algorithm selects a set $F$ of message fields in $\mathit{msg}$ that appear in the condition of $r$ (line 3). Using an SMT solver~\cite{Clarke2018}, the algorithm solves the condition of $r$ to find a set $F^\prime$ of message fields that satisfy the condition (line 4). Specifically, we use Z3~\cite{DeMoura2008:Z3} -- a well-known and widely used SMT solver -- to solve such conditions. Line 5 of the algorithm then replaces the original fields $F$ with the computed fields $F^\prime$. For example, when \FRAM fuzzes a control message guided by the condition $\mathit{version}$ $>$ $5$ $\wedge$ $\mathit{length}$ $\geq$ $10$, it assigns 6 to the $\mathit{version}$ field and 20 to the $\mathit{length}$ field of the control message as the assignments satisfy the condition.

Algorithm~\ref{alg:guided fuzzing} modifies a single control message $\mathit{msg}$ and outputs one fuzzed message $\mathit{msg}^\prime$. Hence, the fuzzing step executes the algorithm $n$ times to generate $n$ number of fuzzed control messages. As shown on lines 6-10, the algorithm updates the budget distribution $B$ with $(r,b{-}1)$ indicating that the rule $r$ has been applied once. Note that the fuzzing step reruns the system under test for each execution of the algorithm.

%\textbf{Mutation.}
As shown on lines 11-17 of Algorithm~\ref{alg:guided fuzzing}, the fuzzing step leverages a mutation technique to diversify fuzzed control messages. Recall that lines 1-5 of the algorithm modify only the message fields that appear in decision rules. Without mutation, decision rules inferred in the first iteration of \FRAM would determine the message fields being modified in all subsequent iterations, while other message fields would remain unchanged. Such a fuzzing method might miss important failure-inducing rules related to other unchanged fields.

The fuzzing step employs a uniform mutation operator~\cite{Talbi2009} that randomly selects fields in a control message with a mutation rate $\mathit{mu}$ and changes the fields' values to random values within their ranges. As shown on line 12 of Algorithm~\ref{alg:guided fuzzing}, the fuzzing step selects the fields in the sniffed control message $\mathit{msg}$ that are not present in the exploited rule $r$. Hence, the return message $\mathit{msg}^\prime$ (line 14) also satisfies the condition of $r$, as mutated fields cannot affect it.
For example, suppose a packet\_in message encoded in 30 fields is sniffed to be fuzzed by \FRAM, and its version and length fields are included in the condition $\mathit{version} > 5 \wedge \mathit{length} \geq 10$ and hence fuzzed (lines 1-10). In this setting, \FRAM randomly selects fields (e.g., $ \mathit{reason}$ and $ \mathit{table\_id}$) that are not present in the condition, and then mutates the selected fields by assigning new random values within their ranges (lines 11-16).
%(e.g., $\mathit{reason} = 8$ and $\mathit{table\_id} = 5673$)

% <<<<< Revision 2
We note that the computational complexity of Algorithm~\ref{alg:guided fuzzing} is primarily determined by line~4, which uses Z3.
The remaining computations scale linearly with the number of message fields.
In our context, as described in Section~\ref{subsec:learning step}, the rule $r$ to be solved by Z3 is concise.
Therefore, Algorithm~\ref{alg:guided fuzzing} is expected to run in practical time, as empirically evaluated in our experiments (Section~\ref{subsec:results}).
% ===== Revision 2

% !TEX root =  ../paper.tex
\section{Evaluation}
\label{sec:evaluation}

In this section, we present our empirical evaluation of \FRAM. Our full evaluation package is available online~\cite{Artifacts}.

\subsection{Research Questions}
\label{subsec:rq}

\noindent\textbf{RQ1 (comparison):}
\textit{How does \FRAM perform compared with state-of-the-art testing techniques for SDNs?}
We investigate whether \FRAM can outperform existing techniques: DELTA~\cite{Lee2017:DELTA} and BEADS~\cite{Jero2017:BEADS} described in Section~\ref{subsec:setup}. We choose these techniques as they rely on fuzzing to test SDN-systems and their implementations are available online. Note that none of the prior methods that identify failure-inducing inputs~\cite{Gopinath2020,KampmannHSZ20} account for the specificities of SDNs; hence, they are not applicable.

\noindent\textbf{RQ2 (usefulness):}
\textit{Can \FRAM learn failure-inducing models that accurately characterize conditions under which a system fails?}
We investigate whether or not \FRAM can infer accurate failure-inducing models. In addition, we compare the failure-inducing conditions identified by \FRAM with those reported in the literature~\cite{Jero2017:BEADS} to assess if these conditions are consistent with analyses from experts.

\noindent\textbf{RQ3 (scalability):}
\textit{Can \FRAM fuzz control messages and learn failure-inducing models in practical time?}
We analyze the relationship between the execution time of \FRAM and network size. To do so, we conduct experiments with systems of various network sizes.

\subsection{Simulation Platform}
\label{subsec:simulation}

To evaluate \FRAM, we opt to use a simulation platform that emulates physical networks. Specifically, we use Mininet~\cite{Lantz2010:Mininet} to create virtual networks of different sizes. In addition, as Mininet employs real SDN switch programs, the emulated networks are very close to real-world SDNs. Hence, Mininet has been widely used in many SDN studies~\cite{Jero2017:BEADS,Lee2017:DELTA,ShinNSB0Z20}. We note that \FRAM can be applied to test actual SDN-systems. However, using physical networks is prohibitively expensive for performing the types of large experiments involved in our systematic evaluations of \FRAM. We ran all our experiments on 10 virtual machines, each of which with 4 CPUs and 10GB of memory. These experiments took $\approx$45 days by concurrently running them on the 10 virtual machines.

\subsection{Study subjects}
\label{subsec:study subjects}

We evaluate \FRAM by testing two actual SDN controllers, i.e., ONOS~\cite{Berde2014:ONOS} and RYU~\cite{RYU}, which are still maintained actively and have been widely used in both research and practice~\cite{Lee2017:DELTA, Jero2017:BEADS, ShuklaSSCZF20, LeeWKYPS20, LiWYYSWZ19, Zhang17}.
Both controllers are implemented using the OpenFlow SDN protocol specification~\cite{OpenFlowSpec}. Since \FRAM fuzzes OpenFlow control messages, it can test any SDN controller that implements the OpenFlow specification.
Regarding virtual networks, we synthesize five networks with 1, 3, 5, 7, and 9 switches controlled by either ONOS or RYU.
In each network, all switches are interconnected with one another, i.e., fully connected topology.
Each switch is connected to two hosts that can emulate any device that sends and receives data streams, e.g., video and sound streams.
We note that our study subjects, i.e., 5 $\times$ 2 SDN-systems built on the five networks controlled by ONOS and RYU, are representative of existing SDN studies and real-world SDNs.
For example, DELTA~\cite{Lee2017:DELTA} (resp. BEADS~\cite{Jero2017:BEADS}) was evaluated with an SDN-system including two (resp. three) switches controlled by ONOS and RYU, as running experiments with SDNs requires large computational resources~\cite{Bannour2018}.
\citet{ShinNSB0Z20} introduced an industrial SDN-system developed in collaboration with SES, a satellite operator, which contains seven switches controlled by ONOS.

\subsection{Experimental setup}
\label{subsec:setup}

\noindent\textbf{EXP1.}
To answer RQ1, we compare \FRAM with DELTA~\cite{Lee2017:DELTA} and BEADS~\cite{Jero2017:BEADS}, which are applicable to our study subjects. DELTA is a security assessment framework for SDNs that enables engineers to automatically reproduce known SDN-related attack scenarios and discover new attack scenarios. For the latter, DELTA relies on random fuzzing that randomizes all fields of a control message without accounting for the specifics of the OpenFlow protocol. BEADS is an automated attack discovery technique based on fuzzing that assumes the OpenFlow protocol, aiming at generating fuzzed control messages that can pass beyond the message parsing layer of the system under test. We note that we reused the implementations available online. However, we had to adapt them in order to make them work in our experiments, though we minimized changes, since the original executables of DELTA and BEADS did not work even after discussions with the authors.

In EXP1, we use two synthetic systems with one switch controlled by either ONOS or RYU. For the test procedure (see Section~\ref{sec:approach}) in EXP1, we use the pairwise ping test~\cite{RFC1122:PING}, applied in many SDN studies~\cite{Lee2017:DELTA, Jero2017:BEADS, Dhawan2015:SPHINX, Canini2012:NICE}, that allows us to detect whether or not hosts can communicate with one another. Regarding the failure detection mechanism (see Section~\ref{sec:approach}) in EXP1, we consider switch disconnections as system failures since both DELTA and BEADS analyzed switch disconnections in their experiments. We further note that EXP1 identifies switch disconnections as failures only when they lead to a communication breakdown and the system fails to locate the causes of the failures, i.e., no relevant log messages related to the failures. EXP1 fuzzes the packet\_in message~\cite{OpenFlowSpec}, which has 57 bytes encoded in 30 fields, as SDN switches send this message to the controller in the execution of the test procedure, and both DELTA and BEADS fuzz it. For details of OpenFlow messages, we refer readers to the OpenFlow specification~\cite{OpenFlowSpec}. We compare the number of fuzzed control messages that cause the switch disconnection failure across fuzzing approaches.

\noindent\textbf{EXP2.}
To answer RQ2, we evaluate the accuracy of failure-inducing models inferred by \FRAM. To this end, we compare the models obtained by \FRAM with those produced by our baselines extending DELTA and BEADS. In addition, we examine our failure-inducing models in light of the literature~\cite{Lee2017:DELTA,Jero2017:BEADS} discussing failure-inducing conditions.

\emph{EXP2.1.} As baselines, we extend DELTA and BEADS, named DELTA$^L$ and BEADS$^L$, to produce failure-inducing models. DELTA$^L$ (resp. BEADS$^L$) encodes the fuzzing results obtained by DELTA (resp. BEADS$^L$) as a training dataset (see the dataset format described in Section~\ref{subsec:fuzzing step:initial}) and uses RIPPER to learn a failure-inducing model from it. Unlike \FRAM, DELTA$^L$ and BEADS$^L$ do not leverage the inferred failure-inducing models to guide their fuzzing.

For ensuring fair comparisons of \FRAM, DELTA$^L$, and BEADS$^L$, EXP2.1 creates a test dataset containing 5000 fuzzing results for each method. EXP2.1 then measures the accuracy of the failure-inducing models obtained by the three methods using the standard precision and recall metrics~\cite{WittenFH16}.
% <<<<< Revision 1
Additionally, EXP2.1 measures imbalance ratios of the datasets obtained at each iteration of the three methods using the imbalance ratio metric~\cite{WittenFH16}.
% ===== Revision 1

We compute precision and recall values as follows: (1)~precision $P$ $=$ $\mathit{TP}/(\mathit{TP}+\mathit{FP})$ and (2)~recall $R$ $=$ $\mathit{TP}/(\mathit{TP}+\mathit{FN})$, where $\mathit{TP}$, $\mathit{FP}$, and $\mathit{FN}$ denote the number of true positives, false positives, and false negatives, respectively. A true positive is a control message labeled with $\mathit{presence}$ (see Section~\ref{subsec:fuzzing step:initial}) and correctly classified as such. A false positive is a control message labeled with $\mathit{absence}$ (see Section~\ref{subsec:fuzzing step:initial}) but incorrectly classified as $\mathit{presence}$. A false negative is a control message labeled with $\mathit{presence}$ but incorrectly classified as $\mathit{absence}$.
% <<<<< Revision 1
We compute the imbalance ratio of a dataset as follows: imbalance ratio $I = 1 - (\mathit{minor} / \mathit{major})$, where $\mathit{minor}$ and $\mathit{major}$ denote the number of control messages in the dataset $D$, labeled with the minority and majority class, respectively.
% ===== Revision 1
In EXP2.1, we use two synthetic systems with one switch controlled by either ONOS or RYU. EXP2.1 applies the pairwise ping test and fuzzes 8000 packet\_in messages with \FRAM, DELTA$^L$, and BEADS$^L$.
%We repeat EXP2.1 10 times.

\emph{EXP2.2.} \citet{Jero2017:BEADS} manually inspected their SDN testing results obtained with BEADS and identified some conditions on message fields that led to system failures. EXP2.2 examines our failure-inducing models to assess the extent to which our models are consistent with their manual analysis results.

For EXP2.2, we use two synthetic systems with one switch controlled by either ONOS or RYU.
% <<<<< Revision 1
EXP2.2 fuzzes the following five types of control messages: packet\_in (57 bytes, 30 fields), hello (8 bytes, 4 fields), barrier\_reply (8 bytes, 4 fields), barrier\_request (8 bytes, 4 fields), and flow\_removed (55 bytes, 22 fields), which are manipulated by BEADS.
We randomly selected these five message types from the 16 types of control messages analyzed in the prior study of BEADS, to keep the expected cost of running our experiments manageable (see Section~\ref{subsec:param}).
% ===== Revision 1
In EXP2.2, \FRAM fuzzes 8000 control messages of each type.
% <<<<< Revision 1
To fuzz the barrier\_request and the barrier\_reply control messages, we use a test procedure that connects switches to an SDN controller as it generates the control messages.
For the remaining types of control messages, we use the pairwise ping test~\cite{RFC1122:PING}.
Regarding failure types, EXP2.2 detects unexpected broadcasts from switches and unexpected switch disconnections as failures.
These are studied in existing work (BEADS).
The broadcasting mechanism of ARP (Address Resolution Protocol)~\cite{ARPSpec} is used to discover host locations in the SDN.
If broadcasting occurs unexpectedly, it may lead to the installation of incorrect forwarding rules on the switches in the SDN, possibly resulting in information leakage (since data can be forwarded to unintended hosts) or connectivity losses.
Regarding the other types of failure, if the communication between the controller and the switches is unexpectedly disconnected, the SDN-system obviously cannot operate as intended.
This is because the controller monitors the network status based on the messages received from the switches and the behavior of the switches is directed by the controller.
% ===== Revision 1

\noindent\textbf{EXP3.}
To answer RQ3, we study the correlation between the execution time of \FRAM and the 2 $\times$ 5 synthetic systems (described in Section~\ref{subsec:study subjects}) with 1, 3, 5, 7, and 9 switches controlled by either ONOS or RYU, respectively. We measure the execution time of each iteration of \FRAM and the execution time of configuring Mininet and the SDN controllers. Our conjecture is that the execution time of \FRAM does not depend on network sizes. However, we also conjecture that there is a correlation between the configuration time of Mininet and SDN controllers and network sizes. Such configuration time includes the time for initializing controllers and Mininet, creating virtual networks, and activating controllers. EXP3 uses the pairwise ping test.

\subsection{Parameter Tuning}
\label{subsec:param}

Recall from Section~\ref{sec:approach} that \FRAM must be configured with the following parameters: number of control messages to be fuzzed at each iteration, mutation rate, and RIPPER parameters.
For tuning the parameters, we ran initial experiments relying on hyperparameter optimization~\cite{WittenFH16} based on guidelines in the literature~\cite{WittenFH16,Hutter2019}.
In our initial experiments, we assessed 10 configurations of the parameters' values to select the best one to be used in further experiments. We selected these 10 configurations using a grid search~\cite{WittenFH16}.  To select the best configuration, for each configuration, we ran \FRAM for four days to ensure there were no notable changes in the results and measured the precision and recall values of the obtained failure-inducing model. For our experiments, we set the number of control messages to be fuzzed at each iteration to 200 and the mutation rate to $1/|F|$, where $|F|$ denotes the number of fields in a control message to be fuzzed. The parameter values of RIPPER used in our experiments can be found in our repository~\cite{Artifacts}.

To fairly compare \FRAM and the other approaches (i.e., DELTA, BEADS, DELTA$^L$, BEADS$^L$), we assign to them the same computation budget: four days for ONOS and two days for RYU.
Within this budget, \FRAM generates a balanced dataset (described in Section~\ref{subsec:learning step}), and precision and recall values of the inferred failure-inducing models reach their plateaus. We note that the configuration time of RYU to run \FRAM is approximately half that of ONOS. Hence, we set different budgets so that \FRAM fuzzes similar numbers of control messages for ONOS and RYU. Since \FRAM is randomized, we repeat our experiments 10 times.
% \textcolor{blue}{Such time budgets are acceptable in practice as \FRAM can be executed offline without engineers' interventions. For example, the BEADS study required eight days to discover failure-inducing control messages.}

The parameters of \FRAM and our experiments can certainly be further tuned to improve the accuracy of \FRAM. However, we were able to convincingly and clearly support our conclusions with the selected configuration, using the study subjects (described in Section~\ref{subsec:study subjects}). Hence, this article does not report further experiments on optimizing those parameters.

%RIPPER: number of folds (2,3(d),4,5), minimum total weights of rules (1,2(d),3,4,5), number of optimization runs (1,2(d),3,4,5), pruning (true), check the error rate (true).
%mutation rate = 0/(field size), 0.5/(filed size), 1 / (field size)
%the number of samples = 200, 400, 600
%After HPO: number of folds = 5, minimum total weights of ruls = 5, number of optimization rules = 2, mutation rate = 1/(field size), the number of samples = 200.
%precision 0.6 and recall = 0.9
%ONOS

\subsection{Experiment Results}
\label{subsec:results}

\begin{figure}[t]
	\centering
	\includegraphics[width=0.65\linewidth]{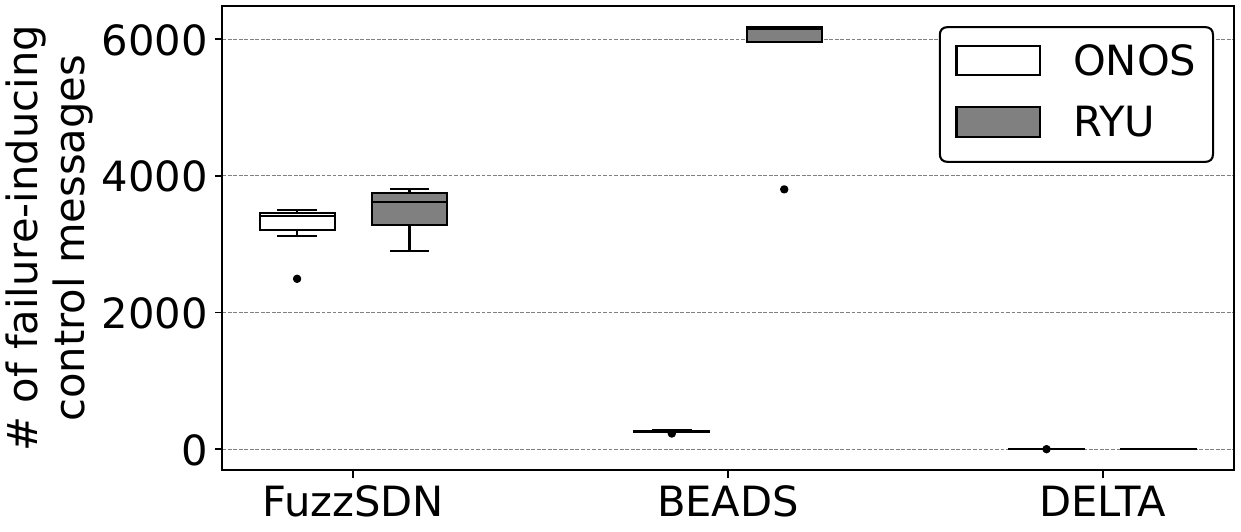}
	\caption{Comparing \FRAM, BEADS, and DELTA based on the number of fuzzed control messages that cause the switch disconnection failure. The boxplots (25\%-50\%-75\%) show distributions of the numbers of failure-inducing control messages obtained from 10 runs of EXP1, testing either ONOS or RYU.}
	\label{fig:RQ1 fail cout}
%\vspace{-0.5em}
\end{figure}

\noindent\textbf{RQ1.}
Figure~\ref{fig:RQ1 fail cout} compares \FRAM, BEADS and DELTA when testing the two study subjects controlled by either ONOS or RYU (EXP1). The boxplots depict distributions (25\%-50\%-75\% quantiles) of the numbers of control messages that cause the switch disconnection failure. The results shown in the figure are obtained from 10 runs of EXP1. To fairly compare \FRAM, BEADS, and DELTA, they were assigned the same computation budget: four days for ONOS and two days for RYU (as described in Section~\ref{subsec:param}).

As shown in Figure~\ref{fig:RQ1 fail cout}, for ONOS, \FRAM generates significantly more failure-inducing control messages than BEADS and DELTA. On average, 3425 failure-inducing control messages are generated by \FRAM, in contrast to 270 by BEADS and 3 by DELTA. Most of the control messages manipulated by DELTA are filtered out by the message parsing layers of the controllers, which is consistent with the finding reported in the BEADS study~\cite{Jero2017:BEADS}.
Regarding the application of BEADS to RYU, the situation is apparently more complicated. RYU is in fact much less robust than ONOS when handling fuzzed control messages. It is therefore easy to fuzz messages leading to failures with RYU. But recall that \FRAM aims at generating a balanced labeled dataset containing control messages that are associated with both the $\mathit{presence}$ and $\mathit{absence}$ of failures in similar proportions (Section~\ref{subsec:learning step}). This is not the case of BEADS which then generates a very large proportion of failure-inducing control messages with RYU, more than that observed with \FRAM.

\begin{tcolorbox}[enhanced jigsaw,left=2pt,right=2pt,top=0pt,bottom=0pt]
\emph{The answer to} \textbf{RQ1} \emph{is that} \FRAM significantly outperforms BEADS and DELTA. In particular, our experiments show that \FRAM is able to generate a much larger number of control messages that cause failures when the SDN controller is relatively robust to fuzzed messages (e.g., ONOS). Such robustness is a desirable and common feature in industrial SDN controllers.
\end{tcolorbox}

\begin{figure*}[t]
	\centering
    \includegraphics[width=\textwidth]{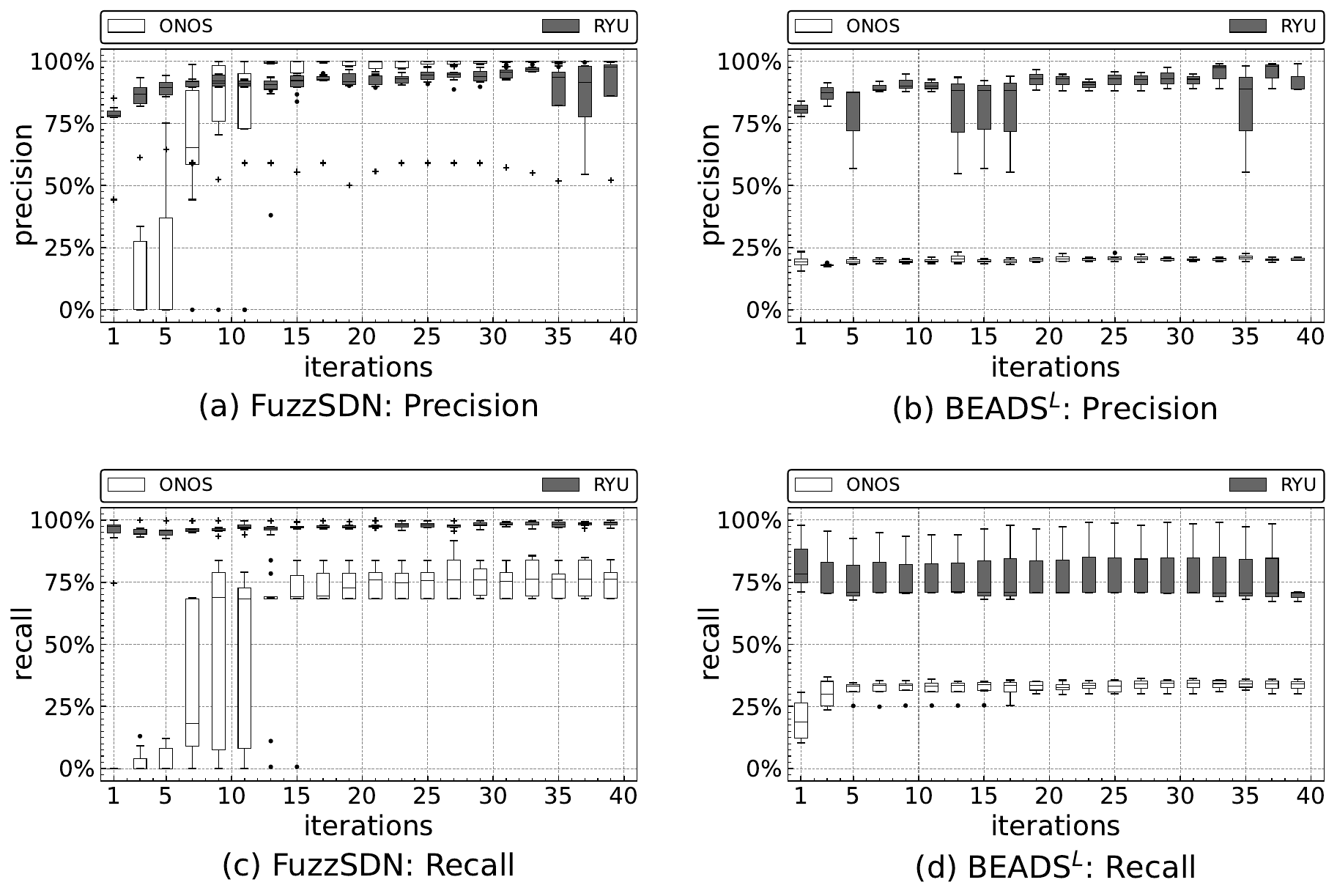}
	\caption{Comparing distributions of precision and recall values obtained from \FRAM and BEADS$^L$ that test the systems controlled by either ONOS or RYU (see EXP2.1). The boxplots (25\%-50\%-75\%) show distributions of precision (a, b) and recall (c, d) values obtained from 10 runs of EXP2.1.}
	\label{fig:exp2 comparison graphs}
%\vspace{-0.5em}
\end{figure*}

\noindent\textbf{RQ2.}
Figure~\ref{fig:exp2 comparison graphs} compares precision (a, b) and recall (c, d) values obtained from \FRAM and BEADS$^L$ for the study subjects controlled by either ONOS or RYU and the test dataset (EXP2.1). The boxplots in Figures~\ref{fig:exp2 comparison graphs}(a) and \ref{fig:exp2 comparison graphs}(b) (resp. \ref{fig:exp2 comparison graphs}(c) and \ref{fig:exp2 comparison graphs}(d)) show distributions (25\%-50\%-75\% quantiles) of precision (resp. recall) values over 40 iterations of the methods obtained from 10 runs of EXP2.1. Note that each iteration of BEADS$^L$ adds 200 fuzzed control messages (the same as \FRAM) to a dataset and learns a failure-inducing model. In contrast to \FRAM, BEADS$^L$ does not use the failure-inducing model to guide fuzzing. We omit the results obtained by DELTA$^L$ because the labeled dataset it created contains only a few failure-inducing control messages, as reported in RQ1.

As shown in Figure~\ref{fig:exp2 comparison graphs}, the failure-inducing models obtained by \FRAM yield higher precision and recall than those obtained by BEADS$^L$ over 40 iterations. The results show that, after 20 iterations, there are no notable changes in precision and recall values. Specifically, for ONOS (resp. RYU), \FRAM achieves, on average, a precision of  99.8\% (resp. 95.4\%) and a recall of 75.5\% (resp. 96.7\%) after 20 iterations. In contrast, BEADS$^L$ achieves, for ONOS (resp. RYU), on average, a precision of 20.9\% (resp. 90.5\%) and a recall of 29.9\% (resp. 70.7\%) after 20 iterations. The 20 iterations of \FRAM took, on average, 2.33 days for ONOS and 1.10 days for RYU.
%which are acceptable in practice as an offline analysis method.

\begin{figure*}[t]
	\centering
    \includegraphics[width=\textwidth]{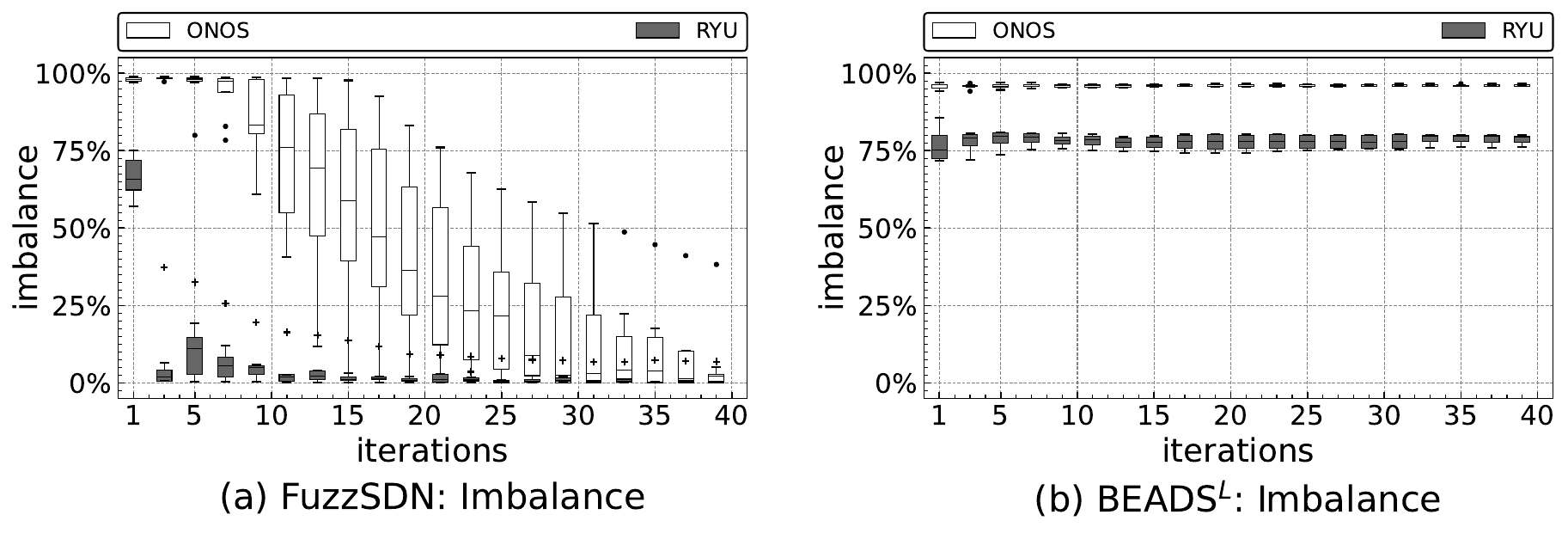}
	\caption{Comparing distributions of imbalance ratios obtained from \FRAM and BEADS$^L$ that test the synthetic systems controlled by either ONOS or RYU (see EXP2.1). The boxplots (25\%-50\%-75\%) show distributions of imbalance ratios obtained from 10 runs of EXP2.1.}
	\label{fig:exp2 imbalance}
%\vspace{-0.5em}
\end{figure*}

% <<<<<< Revision 1
Figure~\ref{fig:exp2 imbalance} shows the comparison of imbalance ratios for datasets obtained from \FRAM and BEADS$^L$ that test the study subjects controlled by either ONOS or RYU (EXP2.1).
The boxplots depict distributions of imbalance ratios over 40 iterations of the methods, compiled from 10 runs of EXP2.1.
Note that the lower the imbalance ratio, the more balanced the dataset.
In Figure~\ref{fig:exp2 imbalance}, we omitted the results obtained by DELTA$^L$ as they provide no additional findings, which are discussed below.
% ===== Revision 1

% <<<<< Revision 1
As shown in Figure~\ref{fig:exp2 imbalance}, the datasets generated by \FRAM are significantly more balanced than those generated by BEADS$^L$ over 40 iterations.
Specifically, after 40 iterations, \FRAM achieves, on average, an imbalance ratio of 5.47\% for ONOS and 1.38\% for RYU.
In contrast, BEADS$^L$ achieves, on average, an imbalance ratio of 96.2\% for ONOS and 78.9\% for RYU.
DELTA$^L$ produces datasets that are even more imbalanced than the other two methods, with, on average, an imbalance ratio of 99.9\% for both ONOS and RYU over 40 iterations.
The results thus indicate that \FRAM effectively addresses the imbalance problem over iterations, leading to accurate characterization of failure-inducing models (see the precision and recall results in Figure~\ref{fig:exp2 comparison graphs}).
However, BEADS$^L$ and DELTA$^L$, which do not tackle the imbalance problem, produce highly imbalanced datasets across all iterations, leading to lower precision and recall of failure-inducing models (see Figure~\ref{fig:exp2 comparison graphs}).
% ===== Revision 1

\begin{table}[t]
\caption{Summary of the EXP2.2 results. Five types of control messages are fuzzed for each experiment with the system controlled by ONOS.}
\label{tbl:RQ2}
\centering
%\footnotesize
\begin{tabularx}{\columnwidth}{@{}Y@{}Y@{}Y@{}Y@{}Y@{}Y@{}}
	\toprule
	message type & message size & \# rules (\FRAM)& \# fields (\FRAM) & \# fields (manual) & all included? \\
	\midrule
	% packet\_in & 57b,30f & 18 & 8 & 3 & yes \\
        packet\_in & 57b,30f & 32 & 12 & 3 & yes \\
	hello & 8b,4f & 21 & 4 & 3 & yes \\
	flow\_removed & 55b,22f & 12 & 4 & 3 & yes \\
	barrier\_request & 8b,4f & 8 & 4 & 3 & yes \\
	barrier\_reply & 8b,4f & 3 & 3 & 3 & yes \\
	\bottomrule
\end{tabularx}

\vspace{0.5em}
\raggedright b: bytes, f: fields
%\vspace{-2.0em}
\end{table}

Table~\ref{tbl:RQ2} presents the summary of our experiment results obtained from EXP2.2. In the experiments, recall that \FRAM fuzzes the following five types of control messages: packet\_in, hello, flow\_removed, barrier\_request, and barrier\_reply. For each control message type, the table shows the message size, the number of rules generated by \FRAM, the number of fields in the inferred rules, and the number of fields that appear in failure-inducing strategies reported in a prior study~\cite{Jero2017:BEADS}. Such strategies were manually defined by analysts, e.g., the switch disconnection failure can occur when changing the version, type, and length fields in a packet\_in message. The ``all included?'' column in the table indicates whether or not the fields in the existing failure-inducing strategies appear in the failure-inducing model obtained by \FRAM.

From Table~\ref{tbl:RQ2}, we see that, for each failure case, all fields in the corresponding failure-inducing strategy described in existing work appear in the failure-inducing models obtained by \FRAM. Hence, the results show that \FRAM does not miss any important field related to system failures. However, \FRAM discovers more fields relevant to failures than the ones reported in prior work. For example, \FRAM found 32 rules with 12 fields for the packet\_in experiment. Nevertheless, we believe that inspecting those 32 precise rules is considerably more efficient for understanding failure-inducing conditions than manually inspecting 8000 fuzzed control messages. Regarding our results for RYU, we refer the reader to our repository~\cite{Artifacts} since our findings are similar to those of ONOS.

\begin{tcolorbox}[enhanced jigsaw,left=2pt,right=2pt,top=0pt,bottom=0pt]
\emph{The answer to} \textbf{RQ2} \emph{is that} \FRAM generates accurate failure-inducing models in practical time, thus faithfully characterizing failure conditions. In addition, the failure analysis results reported in the literature are consistent with the failure-inducing models produced by \FRAM.
\end{tcolorbox}

\begin{figure}[t]
	\centering
	\includegraphics[width=0.65\linewidth]{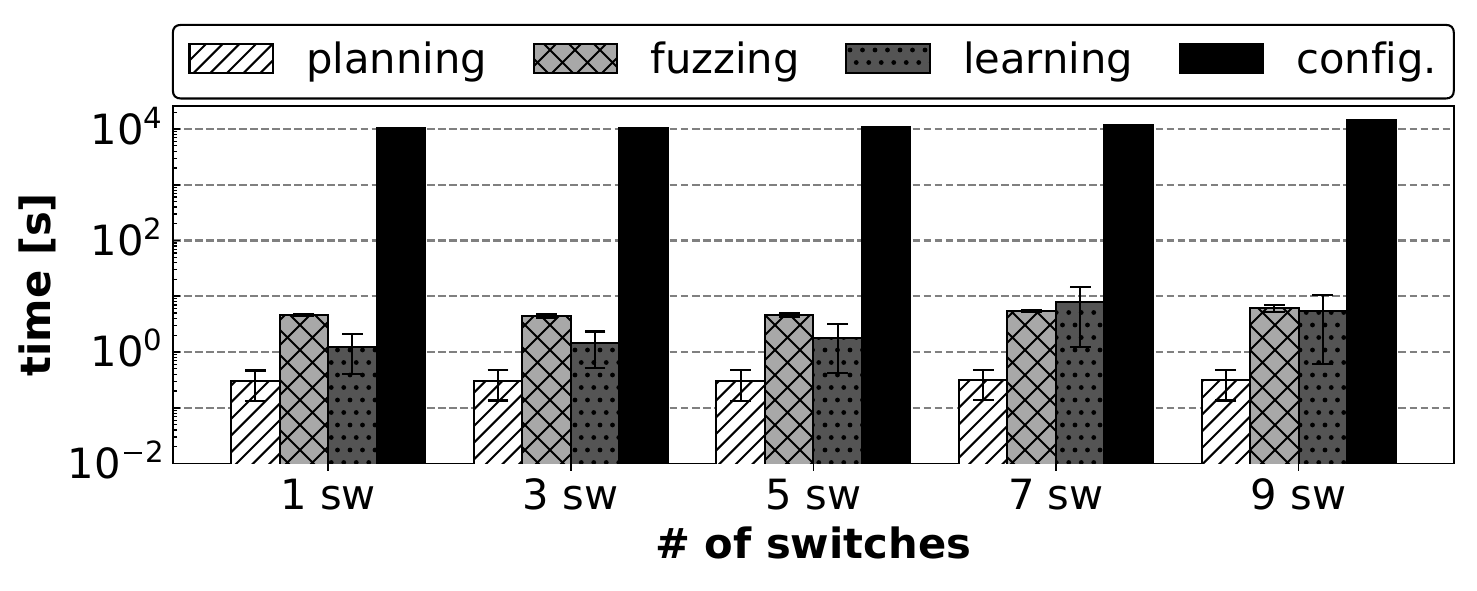}
	\caption{Comparing execution times of \FRAM when varying network sizes as follows: 1, 3, 5, 7, and 9 switches (see EXP3). The bars show the mean execution times of each step of \FRAM and the mean configuration times of ONOS and Mininet computed based on 40 iterations of \FRAM. The vertical lines on the bars show the standard errors of the mean values.}
	\label{fig:RQ3}
%\vspace{-1.5em}
\end{figure}

\noindent\textbf{RQ3.}
Figure~\ref{fig:RQ3} shows the mean execution times of the planning, fuzzing, and learning steps of \FRAM and the mean configuration times of Mininet and ONOS for each iteration of \FRAM. The figure presents the results obtained from EXP3 using five study subjects with 1, 3, 5, 7, and 9 switches controlled by ONOS. Note that the y-axis of the figure is a 10-base log scale in seconds.

As shown in Figure~\ref{fig:RQ3}, the execution times of the three \FRAM steps do not depend on the network sizes.
% <<<<< Revision 2
For each \FRAM iteration that generates, on average, 200 fuzzed control messages, the planning step takes 300ms, the fuzzing step 5.1s, and the learning step 3.5s, which align with our expectations for executing them in practical time.
% ===== Revision 2
However, 200 configurations of Mininet and ONOS at each iteration takes, on average, 2.85h with 1 switch, 3.00h with 3 switches, 3.02h with 5 switches, 3.31h with 7 switches, and 4.07h with 9 switches, which dominate the overall testing times. Recall that the configuration time of RYU takes approximately half that of ONOS. Even though the configuration times of an SDN controller and Mininet are not in the scope of our study, the results indicate a technical bottleneck to be further investigated to shorten testing time.
Regarding the memory requirement, our experiments consumed, at most, 1.2GB of memory, including \FRAM, an SDN controller, and the simulation platform. In particular, ONOS used $\approx$1GB of memory and Mininet $\approx$15MB of memory per switch. Hence, \FRAM does not add significant overhead to the current simulation practice for SDNs.

%For each \FRAM iteration that generates 200 fuzzed control messages, on average \textcolor{orange}{(resp. minimum and maximum)}, the planning step takes 300ms \textcolor{orange}{(resp. 16ms and 662ms)}, the fuzzing step 5.1s \textcolor{orange}{(resp. 3.71s and 8.99s)}, and the learning step 3.5s \textcolor{orange}{(resp. 0.13s and 21.9s)}.
%However, 200 configurations of Mininet and ONOS at each iteration takes, on average \textcolor{orange}{(resp. minimum and maximum)}, 2.85h \textcolor{orange}{(resp. 2.8h and 2.91h)} with 1 switch, 3.00h \textcolor{orange}{(resp. 2.88h and 3.05h)} with 3 switches, 3.02h \textcolor{orange}{(resp. 2.99h and 3.12h)} with 5 switches, 3.31h \textcolor{orange}{(resp. 3.24h and 3.35h)} with 7 switches, and 4.07h \textcolor{orange}{(resp. 3.83h and 4.14h)} with 9 switches, which dominate the overall testing times. Recall that the configuration time of RYU takes approximately half that of ONOS. Even though the configuration times of an SDN controller and Mininet are not in the scope of our study, the results indicate a technical bottleneck to be further investigated to shorten testing time.

\begin{tcolorbox}[enhanced jigsaw,left=2pt,right=2pt,top=0pt,bottom=0pt]
\emph{The answer to} \textbf{RQ3} \emph{is that} the execution time of \FRAM has no correlation with the size of network. Hence, \FRAM is applicable to complex systems with large networks.
\end{tcolorbox}

\subsection{Threats to Validity}
\label{subsec:threats}

% For the development of \FRAM, we focused on known issues reported in the literature, and that were possible to monitor within a reasonable time constraint. We purposely choose to study \emph{Switch disconnection without log information} as they were also studied in previous SDN study, BEADS and DELTA.
% In practice, the studied failure depends on the SDN hardware and controller software used by the SUT. Thus, it is highly specific to the project, and such results could not be compared to existing methods.

%\textcolor{orange}{For the purpose of validating \FRAM, it is important to set a fair time budget to evaluate the performances of the approach against other test method designed for SDN-systems \cite{BoehmeCR2021}.
%Hence, we suggest a time budget of two to four days as it was sufficient to observe plateaus in \FRAM results.
%Shorter time budgets (e.g., 12 hours) may not be sufficient to eliminate the biases in the learning and fuzzing steps, thus lowering the accuracy of \FRAM's generated models.
%In practice, engineers should allot the time budget for modeling a failure depending on the time constraints imposed by the project for the reproduction and isolation of the said failure.}

\noindent\textbf{Internal validity.}
To mitigate potential internal threats to validity, our experiments compared \FRAM with two state-of-the-art solutions (DELTA and BEADS) that generate failure-inducing control messages for testing SDN controllers. Though DELTA and BEADS were our best options, they do not generate failure-inducing models. Therefore, we extended them to produce failure-inducing models and support the comparison of \FRAM with these baselines.

\noindent\textbf{External validity.}
The main concern regarding external validity is the possibility that our results may not generalize to different contexts. In our experiments, we applied \FRAM to several SDNs and two actively maintained SDN controllers, i.e., ONOS and RYU.
We ensured that our synthetic SDN-systems, consisting of five networks with 1, 3, 5, 7, and 9 switches, controlled by either ONOS or RYU, are representative of existing SDN studies and real-world SDN-systems (e.g., emergency management systems~\cite{ShinNSB0Z20}). Further, the SDN-systems used in our experiments are more complex and larger than those previously used to assess DELTA and BEADS, which contain at most three switches.
In general, the performance of SDN fuzzing techniques, such as \FRAM, DELTA, and BEADS, is not correlated with the size of SDN, as these techniques sniff and modify a control message passing through the control channel between the SDN switches and the controller. Specifically, sniffing a network interface and fuzzing a network packet that passes through the interface does not depend on the size of an SDN-system.
To evaluate \FRAM in a realistic setting, we used actual SDN controllers and switch software while emulating only the physical networks, including links, host devices, and switch devices.

% <<<<< Revision 1
The prototype implementation of \FRAM supports OpenFlow, which is a de facto standard SDN protocol used in many SDN-systems~\cite{Lee2017:DELTA, Jero2017:BEADS, ShuklaSSCZF20, Canini2012:NICE, LiWYYSWZ19, LeeWKYPS20}.
As a result, we were able to apply \FRAM to actual SDN controllers (i.e., ONOS and RYU) and compare it with existing tools (i.e., DELTA and BEADS), given their support for OpenFlow.
To apply \FRAM to SDN-systems that employ other SDN protocols, such as Cisco OpFlex~\cite{OpFlexSpec} and ForCES~\cite{ForCESSpec}, one must adapt the sniffing and injecting steps of \FRAM (see Figure~\ref{fig:fuzzing flow}) to correctly decode and encode control messages, respectively.
However, this adaptation does not affect the fuzzing, learning, and planning steps of \FRAM.
We therefore expect that, while the adaptation requires engineering effort to update the sniffing and injecting steps, it does not impact \FRAM's performance.
% ===== Revision 1

\FRAM is developed to be generally applicable to any SDN-system. Our evaluation package and the \FRAM tool are available online~\cite{Artifacts} to (1)~facilitate reproducibility of our experiments and (2)~enable researchers and practitioners to use and adapt \FRAM.
% <<<<< Revision 1
Nevertheless, further case studies in other contexts, including industry SDN-systems that employ different SDN protocols, as well as user studies involving practitioners, remain necessary to further investigate the generalizability of our results.
% ===== Revision 1

\noindent\textbf{Limitations.}
\FRAM requires users to provide a test procedure (e.g., pairwise ping test) and select a control message (e.g., packet\_in) to be fuzzed. These design choices enable \FRAM to generate failure-inducing messages and models within reasonable time budgets (e.g., 4 days for ONOS and 2 days for RYU). Automatically exploring possible use scenarios (i.e., test procedures) and sequences of messages, while accounting for state changes in the SDN controller, can help engineers reduce their manual efforts in testing (e.g., providing test procedures and selecting a message to be fuzzed). However, efficiently and effectively exploring such spaces is a hard problem that requires further research. Given its promising results, we believe that \FRAM is nevertheless a practical solution and serves as a solid foundation for researchers and practitioners to further enhance automation in testing SDN controllers.

% !TEX root =  ../paper.tex
\section{Related Work}
\label{sec:related work}

In this section, we discuss related work in the areas of SDN testing, fuzzing, and characterizing failure-inducing inputs.

\noindent\textbf{SDN testing.}
Testing SDNs has been primarily studied in the networking literature targeting various objectives, such as detecting security vulnerabilities and attacks~\cite{NandaZDWY16,BhuniaG17,Jero2017:BEADS,Lee2017:DELTA,Zhang17,Alshanqiti19,ChicaIB20}, identifying inconsistencies among the SDN components (i.e., applications, controllers, and switches)~\cite{LiWYYSWZ19,LeeWKYPS20,ShuklaSSCZF20}, and analyzing SDN executions~\cite{Canini2012:NICE,DurairajanSB14,StoenescuDPNR18}.
% <<<<< Revision 1
Among these, we discuss SDN testing techniques that rely on fuzzing, as they constitute the most closely related research.
\citet{WooLKS18} proposed RE-CHECKER to fuzz RESTful services provided by SDN controllers.
RE-CHECKER fuzzes an input file, encoded in JSON format, that a network administrator uses to specify network policies (e.g., data forwarding rules).
This results in generating numerous malformed REST messages for testing RESTful services in SDN.
\citet{DixitDS0A18} presented AIM-SDN to test the implementation of the network management datastore architecture (NMDA)~\cite{rfc8342} in SDN.
AIM-SDN randomly fuzzes REST messages to test the NMDA implementation in SDN with regard to the availability, integrity, and confidentiality of datastores. 
\citet{ShuklaSSCZF20} developed PAZZ that aims at detecting faults in SDN switches by fuzzing data packet headers, e.g., IPv4 and IPv6 headers.
\citet{AlbabDHKSWTGY22} presented SwitchV to validate the behaviors of SDN switches.
SwitchV uses fuzzing and symbolic execution to analyze the p4~\cite{BosshartDGIMRSTVVW14} models that specify the behaviors of SDN switches.   
\citet{LeeWKYPS20,LeeWKNYPS22} introduced AudiSDN that employs fuzzing to detect policy inconsistencies among SDN components (i.e., controllers and switches).
AudiSDN fuzzes network policies submitted by administrators through the REST APIs.
To increase the likelihood of discovering inconsistencies, AudiSDN employs rule dependency trees derived from the OpenFlow specification, which restrict valid relationships among rule elements.
In contrast to these existing methods, \FRAM fuzzes SDN control messages to test SDN controllers (as DELTA and BEADS do).
Furthermore, \FRAM uses ML to guide fuzzing and learn failure-inducing models.
% ===== Revision 1

\noindent\textbf{Fuzzing.}
To efficiently generate effective test data, fuzzing has been widely applied in many application domains~\cite{Manes2021:ASEF}. The research strands that most closely relate to our work are fuzzing techniques based on ML for testing networked systems~\cite{Chen19,ZhaoLWSH19}. \citet{Chen19} proposed a fuzzing technique for testing cyber-physical systems, which contain sensors and actuators distributed over a network. The proposed approach relies on a deep learning technique to fuzz actuators' commands that can drive the CPS under test into unsafe physical states. \citet{ZhaoLWSH19} developed SeqFuzzer that enables engineers to test communication systems without prior knowledge of the systems' communication protocols. SeqFuzzer infers communication protocols using a deep learning technique and generates test data based on the inferred protocols. Unlike these prior research threads, \FRAM applies ML and fuzzing in the context of testing SDN-systems. In addition, \FRAM employs an interpretable ML technique to provide engineers with comprehensible failure-inducing models.

% <<<<< Revision 1
Besides the fuzzing techniques mentioned above that leverage ML for testing networked systems, coverage-aware fuzzing techniques have been used in many studies across different domains~\cite{Zalewski:AFL,MaierEFH20,PhamBR20}.
For example, AFL~\cite{Zalewski:AFL} is a mutational, coverage-guided fuzzer that uses compile-time instrumentation and genetic algorithms to automatically generate test cases that uncover previously unexplored internal states in the program under test.
AFL++~\cite{MaierEFH20} is a fuzzing framework that expands on AFL, incorporating many state-of-the-art techniques that enhance fuzzing performance, thus enabling researchers to evaluate different combinations of such techniques.
However, applying such fuzzing tools, developed in other contexts, to test SDN controllers is far from being straightforward as there are differences in inputs, outputs, and states.
For example, the notion of coverage, e.g., statements and branches, used in coverage-aware fuzzing tools is suitable for testing stateless programs, where outputs depend solely on inputs and do not depend on any memory of past interactions.
However, an SDN controller is a stateful program. 
It takes as input sequences of control messages from the switches in the network, processes these messages, and outputs appropriate sequences of responses.
Note that the controller's response is determined by the currently received message and its internal state, which is determined by previously processed messages.
AFLNet~\cite{PhamBR20}, which extends AFL to test servers, is proposed to address this issue by utilizing state coverage rooted in finite state machines.
In AFLNet, finite state machines are constructed using the response codes (also known as status codes) from network protocols, which indicate the result of a client's request to a server.
However, AFLNet is not applicable when such response codes are not available, as in our context.
Hence, we need a different notion of coverage when testing SDN controllers.
In addition to the coverage issue, existing techniques that enhance fuzzing performance for single programs are not easily applicable to testing SDN controllers.
For example, the forkserver technique implemented in AFL++, which uses the fork mechanism to reduce the high cost of initialization, is not applicable to our context.
Since an SDN controller interacts with multiple switches connected to various hosts, testing the controller requires the costly initialization of not only the controller but also the other components in the SDN-system.
Further, testing the controller impacts the states of these other components, making it difficult to efficiently reset and maintain a consistent testing environment.
While coverage-aware fuzzing tools, such as AFL++, provide significant advances in testing stateless programs, applying them to test SDN controllers therefore raises difficult challenges.
% ===== Revision 1

\noindent\textbf{Characterizing failure-inducing inputs.} Recently, a few research strands aimed at characterizing input conditions under which a system under test fails~\cite{Gopinath2020,KampmannHSZ20}. \citet{Gopinath2020} introduced DDTEST that abstracts failure-inducing inputs. DDTEST aims at testing software programs, e.g., JavaScript translators and command-line utilities, that take as input strings. For abstracting failure-inducing inputs, DDTEST uses a derivation tree that depicts how failure-inducing strings can be derived. \citet{KampmannHSZ20} presented ALHAZEN that learns circumstances under which the software program under test fails. ALHAZEN also targets software programs that process strings. ALHAZEN relies on ML to learn failure-inducing circumstances in the form of decision trees. Compared to our work, we note that the context of these research threads is significantly different from our application context: SDNs. Hence, they are not applicable to generate test data and learn failure-inducing models for SDN controllers. To our knowledge, \FRAM is the first attempt that applies ML for guiding fuzzing and learning failure-inducing models by accounting for the specificities of SDNs. 

% !TEX root =  ../paper.tex
\section{Conclusions}
\label{sec:conclusion}

We developed \FRAM, an ML-guided fuzzing method for testing SDN-systems. \FRAM employs ML to guide fuzzing in order to (1)~generate a set of test data, i.e., fuzzed control messages, leading to failures and (2)~learn failure-inducing models that describe conditions when the system is likely to fail. \FRAM implements an iterative process that fuzzes control messages, learns failure-inducing models, and plans how to better guide fuzzing in the next iteration based on the learned models. We evaluated \FRAM on several synthetic SDN-systems controlled by either one of two SDN controllers. Our results indicate that \FRAM is able to generate effective test data that cause system failures and produce accurate failure-inducing models.
%In addition, we compared \FRAM with two state-of-the-art test generation techniques (fuzzing SDNs) and two learning baselines (learning failure-inducing models). The results show that \FRAM significantly outperforms them.
Furthermore, \FRAM's performance does not depend on the network size and is hence applicable to systems with large networks.

%\textcolor{orange}{\textbf{Limitations} \FRAM does not account for the stateful nature of SDN protocol and thus do not account for the dependency between messages sent within the control channels of the SDN system.
%Instead, it focus on a specific of message defined according to the test procedure.
%\FRAM is effective when the attack scenario leading to a failure as been identified beforehand (e.g., unexpected packet\_in conditions can cause a switch disconnection), but there is still a manual effort required from the system developer to manually discover such attack scenario and the type of message that lead to the failure.
%Augmenting \FRAM with the capability of accounting for sequences of messages, would support this effort by removing the need for engineers to identify the attack scenario. Instead, only the effort of detecting a failure and creating a detection mechanism would be left to engineer. However, such a method would rise new challenges such as the exponential increase in the search space, or learning models from sequential and dependent data.}

% <<<<< Revision 1
In the future, we plan to extend \FRAM to account for sequences of control messages. Fuzzing multiple control messages at once poses new challenges due to message intervals, message dependencies, and state changes in the system under test.
To learn failure-inducing models from control message sequences, we will further investigate ML techniques that process sequential data.
In addition, we will extend \FRAM to generate diverse test cases, defined by message sequences, aiming at maximizing state coverage during testing of SDN controllers.
To this end, we will conduct further research to define and quantify the diversity of test cases and the state coverage of SDN controllers, and incorporate these into our fuzzing framework.
% ===== Revision 1
%In addition, we would like to incorporate additional analysis capabilities into \FRAM in order to detect feature interaction problems among multiple SDN applications that utilize a shared set of network resources. 
In the long term, we plan to further validate the generalizability and usefulness of \FRAM by applying it to additional SDN-systems and conducting user studies.

% <<<<< Revision 2
\section*{Data Availability}
\label{sec:data availability}
Our evaluation package and the \FRAM tool are available online~\cite{Artifacts} to (1)~increase the reproducibility of our experiments and (2)~enable researchers and practitioners to use and adapt \FRAM.
% ===== Revision 2

% ===== Leftover =====
% \FRAM does not directly address the problem of finding attack scenario that lead to a failure.
% Instead, it addresses the need of identifying the conditions required for reproducing the failure.
% There is still a manual effort required from the system developer to manually identify the type of message that cause the failure.
% \FRAM does not directly address the problem of detecting new bugs that engineers are unaware of. Instead, it addresses the need of reproducing a failure that has been observed at least once. There is still a manual effort needed from the system developer to write a failure detection mechanism that is able to detect such failure.

%\textbf{Future research} The limitation of \FRAM points toward topics to be addressed in future research. Fuzzing multiple control messages at once poses new challenges
% =====

\begin{acks}
This project has received funding from SES, the Luxembourg National Research Fund under the Industrial Partnership Block Grant (IPBG), ref. IPBG19/14016225/INSTRUCT, and NSERC of Canada under the Discovery and CRC programs.
\end{acks}

%\setcitestyle{numbers}
\bibliographystyle{ACM-Reference-Format}
\balance
\bibliography{refs}

%%% -*-BibTeX-*-
%%% Do NOT edit. File created by BibTeX with style
%%% ACM-Reference-Format-Journals [18-Jan-2012].

\begin{thebibliography}{65}

%%% ====================================================================
%%% NOTE TO THE USER: you can override these defaults by providing
%%% customized versions of any of these macros before the \bibliography
%%% command.  Each of them MUST provide its own final punctuation,
%%% except for \shownote{}, \showDOI{}, and \showURL{}.  The latter two
%%% do not use final punctuation, in order to avoid confusing it with
%%% the Web address.
%%%
%%% To suppress output of a particular field, define its macro to expand
%%% to an empty string, or better, \unskip, like this:
%%%
%%% \newcommand{\showDOI}[1]{\unskip}   % LaTeX syntax
%%%
%%% \def \showDOI #1{\unskip}           % plain TeX syntax
%%%
%%% ====================================================================

\ifx \showCODEN    \undefined \def \showCODEN     #1{\unskip}     \fi
\ifx \showDOI      \undefined \def \showDOI       #1{#1}\fi
\ifx \showISBNx    \undefined \def \showISBNx     #1{\unskip}     \fi
\ifx \showISBNxiii \undefined \def \showISBNxiii  #1{\unskip}     \fi
\ifx \showISSN     \undefined \def \showISSN      #1{\unskip}     \fi
\ifx \showLCCN     \undefined \def \showLCCN      #1{\unskip}     \fi
\ifx \shownote     \undefined \def \shownote      #1{#1}          \fi
\ifx \showarticletitle \undefined \def \showarticletitle #1{#1}   \fi
\ifx \showURL      \undefined \def \showURL       {\relax}        \fi
% The following commands are used for tagged output and should be
% invisible to TeX
\providecommand\bibfield[2]{#2}
\providecommand\bibinfo[2]{#2}
\providecommand\natexlab[1]{#1}
\providecommand\showeprint[2][]{arXiv:#2}

\bibitem[Albab et~al\mbox{.}(2022)]%
        {AlbabDHKSWTGY22}
\bibfield{author}{\bibinfo{person}{Kinan~Dak Albab}, \bibinfo{person}{Jonathan
  DiLorenzo}, \bibinfo{person}{Stefan Heule}, \bibinfo{person}{Ali Kheradmand},
  \bibinfo{person}{Steffen Smolka}, \bibinfo{person}{Konstantin Weitz},
  \bibinfo{person}{Muhammad Timarzi}, \bibinfo{person}{Jiaqi Gao}, {and}
  \bibinfo{person}{Minlan Yu}.} \bibinfo{year}{2022}\natexlab{}.
\newblock \showarticletitle{SwitchV: automated {SDN} switch validation with
  {P4} models}. In \bibinfo{booktitle}{\emph{Proceedings of the {ACM} {SIGCOMM}
  2022 Conference}}. \bibinfo{pages}{365--379}.
\newblock


\bibitem[Alshanqiti et~al\mbox{.}(2019)]%
        {Alshanqiti19}
\bibfield{author}{\bibinfo{person}{Abdullah~M. Alshanqiti},
  \bibinfo{person}{Safi Faizullah}, \bibinfo{person}{Sarwan Ali},
  \bibinfo{person}{Maria~Khalid Alvi}, \bibinfo{person}{Muhammad~Asad Khan},
  {and} \bibinfo{person}{Imdadullah Khan}.} \bibinfo{year}{2019}\natexlab{}.
\newblock \showarticletitle{Detecting {DDoS} Attack on {SDN} Due to
  Vulnerabilities in {OpenFlow}}. In \bibinfo{booktitle}{\emph{Proceedings of
  the 2019 International Conference on Advances in the Emerging Computing
  Technologies}}. \bibinfo{pages}{1--6}.
\newblock


\bibitem[Bannour et~al\mbox{.}(2018)]%
        {Bannour2018}
\bibfield{author}{\bibinfo{person}{Fetia Bannour}, \bibinfo{person}{Sami
  Souihi}, {and} \bibinfo{person}{Abdelhamid Mellouk}.}
  \bibinfo{year}{2018}\natexlab{}.
\newblock \showarticletitle{Distributed {SDN} Control: Survey, Taxonomy, and
  Challenges}.
\newblock \bibinfo{journal}{\emph{{IEEE} Communications Surveys \& Tutorials}}
  \bibinfo{volume}{20} (\bibinfo{year}{2018}), \bibinfo{pages}{333--354}.
\newblock


\bibitem[Berde et~al\mbox{.}(2014)]%
        {Berde2014:ONOS}
\bibfield{author}{\bibinfo{person}{Pankaj Berde}, \bibinfo{person}{Matteo
  Gerola}, \bibinfo{person}{Jonathan Hart}, \bibinfo{person}{Yuta Higuchi},
  \bibinfo{person}{Masayoshi Kobayashi}, \bibinfo{person}{Toshio Koide},
  \bibinfo{person}{Bob Lantz}, \bibinfo{person}{Brian O'Connor},
  \bibinfo{person}{Pavlin Radoslavov}, \bibinfo{person}{William Snow}, {and}
  \bibinfo{person}{Guru Parulkar}.} \bibinfo{year}{2014}\natexlab{}.
\newblock \showarticletitle{{ONOS}: Towards an Open, Distributed {SDN} {OS}}.
  In \bibinfo{booktitle}{\emph{Proceedings of the 3rd Workshop on Hot topics in
  Software Defined Networking}}. \bibinfo{pages}{1--6}.
\newblock


\bibitem[Bhunia and Gurusamy(2017)]%
        {BhuniaG17}
\bibfield{author}{\bibinfo{person}{Suman~Sankar Bhunia} {and}
  \bibinfo{person}{Mohan Gurusamy}.} \bibinfo{year}{2017}\natexlab{}.
\newblock \showarticletitle{Dynamic attack detection and mitigation in IoT
  using {SDN}}. In \bibinfo{booktitle}{\emph{Proceedings of the 27th
  International Telecommunication Networks and Applications Conference}}.
  \bibinfo{pages}{1--6}.
\newblock


\bibitem[Björklund et~al\mbox{.}(2018)]%
        {rfc8342}
\bibfield{author}{\bibinfo{person}{Martin Björklund}, \bibinfo{person}{Jürgen
  Schönwälder}, \bibinfo{person}{Philip~A. Shafer}, \bibinfo{person}{Kent
  Watsen}, {and} \bibinfo{person}{Robert Wilton}.}
  \bibinfo{year}{2018}\natexlab{}.
\newblock \bibinfo{title}{{Network Management Datastore Architecture (NMDA)}}.
\newblock \bibinfo{howpublished}{RFC 8342}.
\newblock


\bibitem[Bosshart et~al\mbox{.}(2014)]%
        {BosshartDGIMRSTVVW14}
\bibfield{author}{\bibinfo{person}{Pat Bosshart}, \bibinfo{person}{Dan Daly},
  \bibinfo{person}{Glen Gibb}, \bibinfo{person}{Martin Izzard},
  \bibinfo{person}{Nick McKeown}, \bibinfo{person}{Jennifer Rexford},
  \bibinfo{person}{Cole Schlesinger}, \bibinfo{person}{Dan Talayco},
  \bibinfo{person}{Amin Vahdat}, \bibinfo{person}{George Varghese}, {and}
  \bibinfo{person}{David Walker}.} \bibinfo{year}{2014}\natexlab{}.
\newblock \showarticletitle{{P4:} programming protocol-independent packet
  processors}.
\newblock \bibinfo{journal}{\emph{Computer Communication Review}}
  \bibinfo{volume}{44}, \bibinfo{number}{3} (\bibinfo{year}{2014}),
  \bibinfo{pages}{87--95}.
\newblock


\bibitem[Braden(1989)]%
        {RFC1122:PING}
\bibfield{author}{\bibinfo{person}{Robert~T. Braden}.}
  \bibinfo{year}{1989}\natexlab{}.
\newblock \bibinfo{booktitle}{\emph{{Requirements for Internet Hosts -
  Communication Layers}}}.
\newblock \bibinfo{type}{Information} RFC 1122.
  \bibinfo{institution}{{I}nternet {E}ngineering {T}ask {F}orce ({IETF})}.
\newblock


\bibitem[Brindescu et~al\mbox{.}(2020)]%
        {Brindescu0LS20}
\bibfield{author}{\bibinfo{person}{Caius Brindescu}, \bibinfo{person}{Iftekhar
  Ahmed}, \bibinfo{person}{Rafael Leano}, {and} \bibinfo{person}{Anita Sarma}.}
  \bibinfo{year}{2020}\natexlab{}.
\newblock \showarticletitle{Planning for untangling: Predicting the difficulty
  of merge conflicts}. In \bibinfo{booktitle}{\emph{Proceedings of the 42nd
  International Conference on Software Engineering}}.
  \bibinfo{pages}{801--811}.
\newblock


\bibitem[Chen et~al\mbox{.}(2019)]%
        {Chen19}
\bibfield{author}{\bibinfo{person}{Yuqi Chen}, \bibinfo{person}{Christopher~M.
  Poskitt}, \bibinfo{person}{Jun Sun}, \bibinfo{person}{Sridhar Adepu}, {and}
  \bibinfo{person}{Fan Zhang}.} \bibinfo{year}{2019}\natexlab{}.
\newblock \showarticletitle{Learning-Guided Network Fuzzing for Testing
  Cyber-Physical System Defences}. In \bibinfo{booktitle}{\emph{Proceedings of
  the 34th {IEEE/ACM} International Conference on Automated Software
  Engineering}}. \bibinfo{pages}{962--973}.
\newblock


\bibitem[Chica et~al\mbox{.}(2020)]%
        {ChicaIB20}
\bibfield{author}{\bibinfo{person}{Juan Camilo~Correa Chica},
  \bibinfo{person}{Jenny~Cuatindioy Imbachi}, {and}
  \bibinfo{person}{Juan~Felipe Botero}.} \bibinfo{year}{2020}\natexlab{}.
\newblock \showarticletitle{Security in {SDN:} A comprehensive survey}.
\newblock \bibinfo{journal}{\emph{Journal of Network and Computer
  Applications}}  \bibinfo{volume}{159} (\bibinfo{year}{2020}),
  \bibinfo{pages}{1--23}.
\newblock


\bibitem[Clarke et~al\mbox{.}(2018)]%
        {Clarke2018}
\bibfield{editor}{\bibinfo{person}{Edmund~M. Clarke},
  \bibinfo{person}{Thomas~A. Henzinger}, \bibinfo{person}{Helmut Veith}, {and}
  \bibinfo{person}{Roderick Bloem}} (Eds.). \bibinfo{year}{2018}\natexlab{}.
\newblock \bibinfo{booktitle}{\emph{Handbook of Model Checking}}.
\newblock \bibinfo{publisher}{Springer}.
\newblock


\bibitem[Cohen(1995)]%
        {Cohen1995:FERI}
\bibfield{author}{\bibinfo{person}{William~W. Cohen}.}
  \bibinfo{year}{1995}\natexlab{}.
\newblock \showarticletitle{Fast Effective Rule Induction}. In
  \bibinfo{booktitle}{\emph{Proceedings of the 12th International Conference on
  Machine Learning}}. \bibinfo{pages}{115--123}.
\newblock


\bibitem[Conti et~al\mbox{.}(2016)]%
        {ContiDL16}
\bibfield{author}{\bibinfo{person}{Mauro Conti}, \bibinfo{person}{Nicola
  Dragoni}, {and} \bibinfo{person}{Viktor Lesyk}.}
  \bibinfo{year}{2016}\natexlab{}.
\newblock \showarticletitle{A Survey of Man In The Middle Attacks}.
\newblock \bibinfo{journal}{\emph{{IEEE} Communications Surveys \& Tutorials}}
  \bibinfo{volume}{18}, \bibinfo{number}{3} (\bibinfo{year}{2016}),
  \bibinfo{pages}{2027--2051}.
\newblock


\bibitem[de~Moura and Bj{\o}rner(2008)]%
        {DeMoura2008:Z3}
\bibfield{author}{\bibinfo{person}{Leonardo de Moura} {and}
  \bibinfo{person}{Nikolaj Bj{\o}rner}.} \bibinfo{year}{2008}\natexlab{}.
\newblock \showarticletitle{{Z3}: An Efficient {SMT} Solver}. In
  \bibinfo{booktitle}{\emph{Proceeding of the 14th International Conference on
  Tools and Algorithms for the Construction and Analysis of Systems}}.
  \bibinfo{pages}{337--340}.
\newblock


\bibitem[Dhawan et~al\mbox{.}(2015)]%
        {Dhawan2015:SPHINX}
\bibfield{author}{\bibinfo{person}{Mohan Dhawan}, \bibinfo{person}{Rishabh
  Poddar}, \bibinfo{person}{Kshiteej Mahajan}, {and} \bibinfo{person}{Vijay
  Mann}.} \bibinfo{year}{2015}\natexlab{}.
\newblock \showarticletitle{{SPHINX}: Detecting Security Attacks in
  Software-Defined Networks}. In \bibinfo{booktitle}{\emph{Proceedings of the
  22nd Network and Distributed System Security Symposium}}.
  \bibinfo{pages}{1--16}.
\newblock


\bibitem[Dixit et~al\mbox{.}(2018)]%
        {DixitDS0A18}
\bibfield{author}{\bibinfo{person}{Vaibhav~Hemant Dixit}, \bibinfo{person}{Adam
  Doup{\'{e}}}, \bibinfo{person}{Yan Shoshitaishvili}, \bibinfo{person}{Ziming
  Zhao}, {and} \bibinfo{person}{Gail{-}Joon Ahn}.}
  \bibinfo{year}{2018}\natexlab{}.
\newblock \showarticletitle{{AIM-SDN:} Attacking Information Mismanagement in
  SDN-datastores}. In \bibinfo{booktitle}{\emph{Proceedings of the 2018 {ACM}
  {SIGSAC} Conference on Computer and Communications Security}}.
  \bibinfo{pages}{664--676}.
\newblock


\bibitem[Drutskoy et~al\mbox{.}(2013)]%
        {Drutskoy2013}
\bibfield{author}{\bibinfo{person}{Dmitry Drutskoy}, \bibinfo{person}{Eric
  Keller}, {and} \bibinfo{person}{Jennifer Rexford}.}
  \bibinfo{year}{2013}\natexlab{}.
\newblock \showarticletitle{Scalable Network Virtualization in Software-Defined
  Networks}.
\newblock \bibinfo{journal}{\emph{IEEE Internet Computing}}
  \bibinfo{volume}{17} (\bibinfo{year}{2013}), \bibinfo{pages}{20--27}.
\newblock


\bibitem[Durairajan et~al\mbox{.}(2014)]%
        {DurairajanSB14}
\bibfield{author}{\bibinfo{person}{Ramakrishnan Durairajan},
  \bibinfo{person}{Joel Sommers}, {and} \bibinfo{person}{Paul Barford}.}
  \bibinfo{year}{2014}\natexlab{}.
\newblock \showarticletitle{Controller-agnostic {SDN} Debugging}. In
  \bibinfo{booktitle}{\emph{Proceedings of the 10th {ACM} International on
  Conference on emerging Networking Experiments and Technologies}},
  \bibfield{editor}{\bibinfo{person}{Aruna Seneviratne},
  \bibinfo{person}{Christophe Diot}, \bibinfo{person}{Jim Kurose},
  \bibinfo{person}{Augustin Chaintreau}, {and} \bibinfo{person}{Luigi Rizzo}}
  (Eds.). \bibinfo{pages}{227--234}.
\newblock


\bibitem[Ferr{\'u}s et~al\mbox{.}(2016)]%
        {Ferrus2016}
\bibfield{author}{\bibinfo{person}{Ramon Ferr{\'u}s}, \bibinfo{person}{Harilaos
  Koumaras}, \bibinfo{person}{Oriol Sallent}, \bibinfo{person}{George Agapiou},
  \bibinfo{person}{Tinku Rasheed}, \bibinfo{person}{M-A Kourtis},
  \bibinfo{person}{C Boustie}, \bibinfo{person}{Patrick G{\'e}lard}, {and}
  \bibinfo{person}{Toufik Ahmed}.} \bibinfo{year}{2016}\natexlab{}.
\newblock \showarticletitle{{SDN/NFV}-enabled satellite communications
  networks: Opportunities, scenarios and challenges}.
\newblock \bibinfo{journal}{\emph{Journal of Physical Communication}}
  \bibinfo{volume}{18} (\bibinfo{year}{2016}), \bibinfo{pages}{95--112}.
\newblock


\bibitem[Fioraldi et~al\mbox{.}(2020)]%
        {MaierEFH20}
\bibfield{author}{\bibinfo{person}{Andrea Fioraldi},
  \bibinfo{person}{Dominik~Christian Maier}, \bibinfo{person}{Heiko
  Ei{\ss}feldt}, {and} \bibinfo{person}{Marc Heuse}.}
  \bibinfo{year}{2020}\natexlab{}.
\newblock \showarticletitle{{AFL++} : Combining Incremental Steps of Fuzzing
  Research}. In \bibinfo{booktitle}{\emph{Proceedings of the 14th {USENIX}
  Workshop on Offensive Technologies}}.
\newblock


\bibitem[Ghotra et~al\mbox{.}(2015)]%
        {GhotraMH15}
\bibfield{author}{\bibinfo{person}{Baljinder Ghotra}, \bibinfo{person}{Shane
  McIntosh}, {and} \bibinfo{person}{Ahmed~E. Hassan}.}
  \bibinfo{year}{2015}\natexlab{}.
\newblock \showarticletitle{Revisiting the Impact of Classification Techniques
  on the Performance of Defect Prediction Models}. In
  \bibinfo{booktitle}{\emph{Proceedings of the 37th {IEEE/ACM} International
  Conference on Software Engineering}}. \bibinfo{pages}{789--800}.
\newblock


\bibitem[Godefroid et~al\mbox{.}(2017)]%
        {GodefroidPS17}
\bibfield{author}{\bibinfo{person}{Patrice Godefroid}, \bibinfo{person}{Hila
  Peleg}, {and} \bibinfo{person}{Rishabh Singh}.}
  \bibinfo{year}{2017}\natexlab{}.
\newblock \showarticletitle{Learn{\&}Fuzz: machine learning for input fuzzing}.
  In \bibinfo{booktitle}{\emph{Proceedings of the 32nd {IEEE/ACM} International
  Conference on Automated Software Engineering}}. \bibinfo{pages}{50--59}.
\newblock


\bibitem[Gopinath et~al\mbox{.}(2020)]%
        {Gopinath2020}
\bibfield{author}{\bibinfo{person}{Rahul Gopinath}, \bibinfo{person}{Alexander
  Kampmann}, \bibinfo{person}{Nikolas Havrikov}, \bibinfo{person}{Ezekiel~O.
  Soremekun}, {and} \bibinfo{person}{Andreas Zeller}.}
  \bibinfo{year}{2020}\natexlab{}.
\newblock \showarticletitle{Abstracting failure-inducing inputs}. In
  \bibinfo{booktitle}{\emph{Proceedings of the 29th {ACM SIGSOFT} International
  Symposium on Software Testing and Analysis}}. \bibinfo{publisher}{ACM},
  \bibinfo{pages}{237--248}.
\newblock


\bibitem[Haleplidis et~al\mbox{.}(2015)]%
        {SDN:15}
\bibfield{author}{\bibinfo{person}{Evangelos Haleplidis},
  \bibinfo{person}{Kostas Pentikousis}, \bibinfo{person}{Spyros~G. Denazis},
  \bibinfo{person}{Jamal~Hadi Salim}, \bibinfo{person}{David Meyer}, {and}
  \bibinfo{person}{Odysseas~G. Koufopavlou}.} \bibinfo{year}{2015}\natexlab{}.
\newblock \bibinfo{booktitle}{\emph{Software-Defined Networking ({SDN}): Layers
  and Architecture Terminology}}.
\newblock \bibinfo{type}{Information} RFC 7426. \bibinfo{institution}{Internet
  Research Task Force (IRTF)}.
\newblock


\bibitem[Halpern et~al\mbox{.}(2010)]%
        {ForCESSpec}
\bibfield{author}{\bibinfo{person}{Joel~M. Halpern}, \bibinfo{person}{Robert
  Haas}, \bibinfo{person}{Doria Avri}, \bibinfo{person}{Ligang Dong},
  \bibinfo{person}{Weiming Wang}, \bibinfo{person}{Hormuzd~M. Khosravi},
  \bibinfo{person}{Jamal~Hadi Salim}, {and} \bibinfo{person}{Ram Gopal}.}
  \bibinfo{year}{2010}\natexlab{}.
\newblock \bibinfo{booktitle}{\emph{{Forwarding and Control Element Separation
  (ForCES) Protocol Specification}}}.
\newblock \bibinfo{type}{Information} RFC 5810.
\newblock


\bibitem[Haq et~al\mbox{.}(2021)]%
        {HaqSNB21}
\bibfield{author}{\bibinfo{person}{Fitash~Ul Haq}, \bibinfo{person}{Donghwan
  Shin}, \bibinfo{person}{Shiva Nejati}, {and} \bibinfo{person}{Lionel~C.
  Briand}.} \bibinfo{year}{2021}\natexlab{}.
\newblock \showarticletitle{Can Offline Testing of Deep Neural Networks Replace
  Their Online Testing?}
\newblock \bibinfo{journal}{\emph{Empirical Software Engineering}}
  \bibinfo{volume}{26}, \bibinfo{number}{90} (\bibinfo{year}{2021}),
  \bibinfo{pages}{1--30}.
\newblock


\bibitem[Hutter et~al\mbox{.}(2019)]%
        {Hutter2019}
\bibfield{author}{\bibinfo{person}{Frank Hutter}, \bibinfo{person}{Lars
  Kotthoff}, {and} \bibinfo{person}{Joaquin Vanschoren}.}
  \bibinfo{year}{2019}\natexlab{}.
\newblock \bibinfo{booktitle}{\emph{Automated Machine Learning: Methods,
  Systems, Challenges} (\bibinfo{edition}{1} ed.)}.
\newblock \bibinfo{publisher}{Springer}.
\newblock


\bibitem[Jero et~al\mbox{.}(2017)]%
        {Jero2017:BEADS}
\bibfield{author}{\bibinfo{person}{Samuel Jero}, \bibinfo{person}{Xiangyu Bu},
  \bibinfo{person}{Cristina Nita-Rotaru}, \bibinfo{person}{Hamed Okhravi},
  \bibinfo{person}{Richard Skowyra}, {and} \bibinfo{person}{Sonia Fahmy}.}
  \bibinfo{year}{2017}\natexlab{}.
\newblock \showarticletitle{{BEADS}: Automated Attack Discovery in
  {OpenFlow}-Based {SDN} Systems}. In \bibinfo{booktitle}{\emph{Proceedings of
  the 20th International Symposium on Research in Attacks, Intrusions, and
  Defenses}}. \bibinfo{pages}{311--333}.
\newblock


\bibitem[Kampmann et~al\mbox{.}(2020)]%
        {KampmannHSZ20}
\bibfield{author}{\bibinfo{person}{Alexander Kampmann},
  \bibinfo{person}{Nikolas Havrikov}, \bibinfo{person}{Ezekiel~O. Soremekun},
  {and} \bibinfo{person}{Andreas Zeller}.} \bibinfo{year}{2020}\natexlab{}.
\newblock \showarticletitle{When does my program do this? learning
  circumstances of software behavior}. In \bibinfo{booktitle}{\emph{Proceedings
  of the 28th {ACM} Joint European Software Engineering Conference and
  Symposium on the Foundations of Software Engineering}}.
  \bibinfo{pages}{1228--1239}.
\newblock


\bibitem[Lantz et~al\mbox{.}(2010)]%
        {Lantz2010:Mininet}
\bibfield{author}{\bibinfo{person}{Bob Lantz}, \bibinfo{person}{Brandon
  Heller}, {and} \bibinfo{person}{Nick McKeown}.}
  \bibinfo{year}{2010}\natexlab{}.
\newblock \showarticletitle{A network in a laptop: rapid prototyping for
  software-defined networks}. In \bibinfo{booktitle}{\emph{Proceedings of the
  9th {ACM SIGCOMM} Workshop on Hot Topics in Networks}}.
  \bibinfo{pages}{1--6}.
\newblock


\bibitem[Lara et~al\mbox{.}(2014)]%
        {Lara2014}
\bibfield{author}{\bibinfo{person}{Adrian Lara}, \bibinfo{person}{Anisha
  Kolasani}, {and} \bibinfo{person}{Byrav Ramamurthy}.}
  \bibinfo{year}{2014}\natexlab{}.
\newblock \showarticletitle{Network Innovation using {OpenFlow}: A Survey}.
\newblock \bibinfo{journal}{\emph{{IEEE} Communications Surveys \& Tutorials}}
  \bibinfo{volume}{16} (\bibinfo{year}{2014}), \bibinfo{pages}{493--512}.
\newblock


\bibitem[Lee et~al\mbox{.}(2022)]%
        {LeeWKNYPS22}
\bibfield{author}{\bibinfo{person}{Seungsoo Lee}, \bibinfo{person}{Seungwon
  Woo}, \bibinfo{person}{Jinwoo Kim}, \bibinfo{person}{Jaehyun Nam},
  \bibinfo{person}{Vinod Yegneswaran}, \bibinfo{person}{Phillip~A. Porras},
  {and} \bibinfo{person}{Seungwon Shin}.} \bibinfo{year}{2022}\natexlab{}.
\newblock \showarticletitle{A Framework for Policy Inconsistency Detection in
  Software-Defined Networks}.
\newblock \bibinfo{journal}{\emph{{IEEE/ACM} Transactions on Networking}}
  \bibinfo{volume}{30}, \bibinfo{number}{3} (\bibinfo{year}{2022}),
  \bibinfo{pages}{1410--1423}.
\newblock


\bibitem[Lee et~al\mbox{.}(2020)]%
        {LeeWKYPS20}
\bibfield{author}{\bibinfo{person}{Seungsoo Lee}, \bibinfo{person}{Seungwon
  Woo}, \bibinfo{person}{Jinwoo Kim}, \bibinfo{person}{Vinod Yegneswaran},
  \bibinfo{person}{Phillip~A. Porras}, {and} \bibinfo{person}{Seungwon Shin}.}
  \bibinfo{year}{2020}\natexlab{}.
\newblock \showarticletitle{{AudiSDN}: Automated Detection of Network Policy
  Inconsistencies in Software-Defined Networks}. In
  \bibinfo{booktitle}{\emph{Proceedings of the 39th {IEEE} Conference on
  Computer Communications}}. \bibinfo{pages}{1788--1797}.
\newblock


\bibitem[Lee et~al\mbox{.}(2017)]%
        {Lee2017:DELTA}
\bibfield{author}{\bibinfo{person}{Seungsoo Lee}, \bibinfo{person}{Changhoon
  Yoon}, \bibinfo{person}{Chanhee Lee}, \bibinfo{person}{Seungwon Shin},
  \bibinfo{person}{Vinod Yegneswaran}, {and} \bibinfo{person}{Phillip Porras}.}
  \bibinfo{year}{2017}\natexlab{}.
\newblock \showarticletitle{{DELTA}: A Security Assessment Framework for
  Software-Defined Networks}. In \bibinfo{booktitle}{\emph{Proceedings of the
  24th Network and Distributed System Security Symposium}}.
  \bibinfo{pages}{1--15}.
\newblock


\bibitem[Li et~al\mbox{.}(2019)]%
        {LiWYYSWZ19}
\bibfield{author}{\bibinfo{person}{Yahui Li}, \bibinfo{person}{Zhiliang Wang},
  \bibinfo{person}{Jiangyuan Yao}, \bibinfo{person}{Xia Yin},
  \bibinfo{person}{Xingang Shi}, \bibinfo{person}{Jianping Wu}, {and}
  \bibinfo{person}{Han Zhang}.} \bibinfo{year}{2019}\natexlab{}.
\newblock \showarticletitle{{MSAID:} Automated detection of interference in
  multiple {SDN} applications}.
\newblock \bibinfo{journal}{\emph{Computer Networks}}  \bibinfo{volume}{153}
  (\bibinfo{year}{2019}), \bibinfo{pages}{49--62}.
\newblock


\bibitem[Liu et~al\mbox{.}(2018)]%
        {LiuSZCSK18}
\bibfield{author}{\bibinfo{person}{Jiajia Liu}, \bibinfo{person}{Yongpeng Shi},
  \bibinfo{person}{Lei Zhao}, \bibinfo{person}{Yurui Cao}, \bibinfo{person}{Wen
  Sun}, {and} \bibinfo{person}{Nei Kato}.} \bibinfo{year}{2018}\natexlab{}.
\newblock \showarticletitle{Joint Placement of Controllers and Gateways in
  {SDN}-Enabled {5G}-Satellite Integrated Network}.
\newblock \bibinfo{journal}{\emph{{IEEE} Journal on Selected Areas in
  Communications}} \bibinfo{volume}{36}, \bibinfo{number}{2}
  (\bibinfo{year}{2018}), \bibinfo{pages}{221--232}.
\newblock


\bibitem[Manes et~al\mbox{.}(2021)]%
        {Manes2021:ASEF}
\bibfield{author}{\bibinfo{person}{Valentin~J.M. Manes},
  \bibinfo{person}{HyungSeok Han}, \bibinfo{person}{Choongwoo Han},
  \bibinfo{person}{Sang~Kil Cha}, \bibinfo{person}{Manuel Egele},
  \bibinfo{person}{Edward~J. Schwartz}, {and} \bibinfo{person}{Maverick Woo}.}
  \bibinfo{year}{2021}\natexlab{}.
\newblock \showarticletitle{The Art, Science, and Engineering of Fuzzing: A
  Survey}.
\newblock \bibinfo{journal}{\emph{IEEE Transactions on Software Engineering}}
  \bibinfo{volume}{47} (\bibinfo{year}{2021}), \bibinfo{pages}{2312--2331}.
\newblock
Issue 11.


\bibitem[Marco et~al\mbox{.}(2012)]%
        {Canini2012:NICE}
\bibfield{author}{\bibinfo{person}{Canini Marco}, \bibinfo{person}{Venzano
  Daniele}, \bibinfo{person}{Perešíni Peter}, \bibinfo{person}{Kostić
  Dejan}, {and} \bibinfo{person}{Rexford Jennifer}.}
  \bibinfo{year}{2012}\natexlab{}.
\newblock \showarticletitle{A {NICE} Way to Test {OpenFlow} Applications}. In
  \bibinfo{booktitle}{\emph{Proceedings of the 9th {USENIX} Symposium on
  Networked Systems Design and Implementation}}. \bibinfo{pages}{127--140}.
\newblock


\bibitem[Mirkovic and Reiher(2004)]%
        {MirkovicR2004}
\bibfield{author}{\bibinfo{person}{Jelena Mirkovic} {and}
  \bibinfo{person}{Peter Reiher}.} \bibinfo{year}{2004}\natexlab{}.
\newblock \showarticletitle{A Taxonomy of DDoS Attack and DDoS Defense
  Mechanisms}.
\newblock \bibinfo{journal}{\emph{SIGCOMM Computer Communication Review}}
  \bibinfo{volume}{34}, \bibinfo{number}{2} (\bibinfo{year}{2004}),
  \bibinfo{pages}{39–--53}.
\newblock


\bibitem[Molnar(2022)]%
        {Molnar2022}
\bibfield{author}{\bibinfo{person}{Christoph Molnar}.}
  \bibinfo{year}{2022}\natexlab{}.
\newblock \bibinfo{booktitle}{\emph{Interpretable Machine Learning: A Guide for
  Making Black Box Models Explainable} (\bibinfo{edition}{2} ed.)}.
\newblock
\urldef\tempurl%
\url{https://christophm.github.io/interpretable-ml-book}
\showURL{%
\tempurl}


\bibitem[Moy(1998)]%
        {OSPFSpec}
\bibfield{author}{\bibinfo{person}{John Moy}.} \bibinfo{year}{1998}\natexlab{}.
\newblock \bibinfo{booktitle}{\emph{{OSPF Version 2}}}.
\newblock \bibinfo{type}{Information} RFC 2328. \bibinfo{institution}{Ascend
  Communications, Inc.}
\newblock


\bibitem[Nanda et~al\mbox{.}(2016)]%
        {NandaZDWY16}
\bibfield{author}{\bibinfo{person}{Saurav Nanda}, \bibinfo{person}{Faheem
  Zafari}, \bibinfo{person}{Casimer DeCusatis}, \bibinfo{person}{Eric Wedaa},
  {and} \bibinfo{person}{Baijian Yang}.} \bibinfo{year}{2016}\natexlab{}.
\newblock \showarticletitle{Predicting network attack patterns in {SDN} using
  machine learning approach}. In \bibinfo{booktitle}{\emph{Proceedings of the
  2016 {IEEE} Conference on Network Function Virtualization and Software
  Defined Networks}}. \bibinfo{pages}{167--172}.
\newblock


\bibitem[Ollando et~al\mbox{.}(2023)]%
        {Artifacts}
\bibfield{author}{\bibinfo{person}{Raphael Ollando},
  \bibinfo{person}{Seung~Yeob Shin}, {and} \bibinfo{person}{Lionel~C. Briand}.}
  \bibinfo{year}{2023}\natexlab{}.
\newblock \bibinfo{title}{[Artifact Repository] Learning Failure-Inducing
  Models for Testing Software-Defined Networks}.
\newblock
  \bibinfo{howpublished}{\url{https://figshare.com/s/541ddc973352a8ac193e}}.
\newblock


\bibitem[{Open Networking Foundation}(2015)]%
        {OpenFlowSpec}
\bibfield{author}{\bibinfo{person}{{Open Networking Foundation}}.}
  \bibinfo{year}{2015}\natexlab{}.
\newblock \bibinfo{booktitle}{\emph{{O}pen{F}low Switch Specification, Version
  1.5.1}}.
\newblock \bibinfo{type}{Specification} ONF TS-025. \bibinfo{institution}{Open
  Networking Foundation}.
\newblock


\bibitem[Pham et~al\mbox{.}(2020)]%
        {PhamBR20}
\bibfield{author}{\bibinfo{person}{Van{-}Thuan Pham}, \bibinfo{person}{Marcel
  B{\"{o}}hme}, {and} \bibinfo{person}{Abhik Roychoudhury}.}
  \bibinfo{year}{2020}\natexlab{}.
\newblock \showarticletitle{{AFLNET:} {A} Greybox Fuzzer for Network
  Protocols}. In \bibinfo{booktitle}{\emph{Proceedings of the 13th {IEEE}
  International Conference on Software Testing, Validation and Verification}}.
  \bibinfo{pages}{460--465}.
\newblock


\bibitem[Plummer(1982)]%
        {ARPSpec}
\bibfield{author}{\bibinfo{person}{David~C. Plummer}.}
  \bibinfo{year}{1982}\natexlab{}.
\newblock \bibinfo{booktitle}{\emph{{An Ethernet Address Resolution Protocol:
  Or Converting Network Protocol Addresses to 48.bit Ethernet Address for
  Transmission on Ethernet Hardware}}}.
\newblock \bibinfo{type}{Information}. \bibinfo{institution}{Internet
  Engineering Task Force (IETF)}.
\newblock


\bibitem[Postel(1980)]%
        {UDPSpec}
\bibfield{author}{\bibinfo{person}{Jon Postel}.}
  \bibinfo{year}{1980}\natexlab{}.
\newblock \bibinfo{booktitle}{\emph{User Datagram Protocol}}.
\newblock \bibinfo{type}{Information} RFC 768.
  \bibinfo{institution}{USC/Information Sciences Institute}.
\newblock


\bibitem[Postel(1981)]%
        {TCPSpec}
\bibfield{author}{\bibinfo{person}{Jon Postel}.}
  \bibinfo{year}{1981}\natexlab{}.
\newblock \bibinfo{booktitle}{\emph{Transmission Control Protocol}}.
\newblock \bibinfo{type}{Information} RFC 793.
  \bibinfo{institution}{USC/Information Sciences Institute}.
\newblock


\bibitem[Quinlan(1993)]%
        {Quinlan1993}
\bibfield{author}{\bibinfo{person}{Ross Quinlan}.}
  \bibinfo{year}{1993}\natexlab{}.
\newblock \bibinfo{booktitle}{\emph{{C4.5}: Programs for Machine Learning}}.
\newblock \bibinfo{publisher}{Morgan Kaufmann Publishers}.
\newblock


\bibitem[Rafique et~al\mbox{.}(2020)]%
        {Rafique2020}
\bibfield{author}{\bibinfo{person}{Wajid Rafique}, \bibinfo{person}{Lianyong
  Qi}, \bibinfo{person}{Ibrar Yaqoob}, \bibinfo{person}{Muhammad Imran},
  \bibinfo{person}{Raihan~Ur Rasool}, {and} \bibinfo{person}{Wanchun Dou}.}
  \bibinfo{year}{2020}\natexlab{}.
\newblock \showarticletitle{Complementing {IoT} Services Through Software
  Defined Networking and Edge Computing: A Comprehensive Survey}.
\newblock \bibinfo{journal}{\emph{{IEEE} Communications Surveys \& Tutorials}}
  \bibinfo{volume}{22}, \bibinfo{number}{3} (\bibinfo{year}{2020}),
  \bibinfo{pages}{1761--1804}.
\newblock


\bibitem[Rekhter et~al\mbox{.}(2006)]%
        {BGPSpec}
\bibfield{author}{\bibinfo{person}{Yakov Rekhter}, \bibinfo{person}{Tony Li},
  {and} \bibinfo{person}{Susan Hares}.} \bibinfo{year}{2006}\natexlab{}.
\newblock \bibinfo{booktitle}{\emph{{A Border Gateway Protocol 4 (BGP-4)}}}.
\newblock \bibinfo{type}{Information} RFC 4271. \bibinfo{institution}{Internet
  Engineering Task Force (IETF)}.
\newblock


\bibitem[R\"{o}pke and Holz(2015)]%
        {Ropke2015:Rootkits}
\bibfield{author}{\bibinfo{person}{Christian R\"{o}pke} {and}
  \bibinfo{person}{Thorsten Holz}.} \bibinfo{year}{2015}\natexlab{}.
\newblock \showarticletitle{{SDN} {R}ootkits: Subverting Network Operating
  Systems of Software-Defined Networks}. In
  \bibinfo{booktitle}{\emph{Proceedings of the 18th International Symposium on
  Research in Attacks, Intrusions, and Defenses}}. \bibinfo{pages}{339--356}.
\newblock


\bibitem[{RYU Project Team}(2014)]%
        {RYU}
\bibfield{author}{\bibinfo{person}{{RYU Project Team}}.}
  \bibinfo{year}{2014}\natexlab{}.
\newblock \bibinfo{booktitle}{\emph{{RYU SDN} {F}ramework}
  (\bibinfo{edition}{1} ed.)}.
\newblock \bibinfo{publisher}{RYU Project Team}.
\newblock


\bibitem[Shin et~al\mbox{.}(2020)]%
        {ShinNSB0Z20}
\bibfield{author}{\bibinfo{person}{Seung~Yeob Shin}, \bibinfo{person}{Shiva
  Nejati}, \bibinfo{person}{Mehrdad Sabetzadeh}, \bibinfo{person}{Lionel~C.
  Briand}, \bibinfo{person}{Chetan Arora}, {and} \bibinfo{person}{Frank
  Zimmer}.} \bibinfo{year}{2020}\natexlab{}.
\newblock \showarticletitle{Dynamic adaptation of software-defined networks for
  {IoT} systems: A search-based approach}. In
  \bibinfo{booktitle}{\emph{Proceedings of the 15th {IEEE/ACM} International
  Symposium on Software Engineering for Adaptive and Self-Managing Systems}}.
  \bibinfo{pages}{137--148}.
\newblock


\bibitem[Shukla et~al\mbox{.}(2020)]%
        {ShuklaSSCZF20}
\bibfield{author}{\bibinfo{person}{Apoorv Shukla}, \bibinfo{person}{Said~Jawad
  Saidi}, \bibinfo{person}{Stefan Schmid}, \bibinfo{person}{Marco Canini},
  \bibinfo{person}{Thomas Zinner}, {and} \bibinfo{person}{Anja Feldmann}.}
  \bibinfo{year}{2020}\natexlab{}.
\newblock \showarticletitle{Toward Consistent {SDN}s: A Case for Network State
  Fuzzing}.
\newblock \bibinfo{journal}{\emph{{IEEE} Transactions on Network and Service
  Management}} \bibinfo{volume}{17}, \bibinfo{number}{2}
  (\bibinfo{year}{2020}), \bibinfo{pages}{668--681}.
\newblock


\bibitem[Smith et~al\mbox{.}(2016)]%
        {OpFlexSpec}
\bibfield{author}{\bibinfo{person}{Michael Smith},
  \bibinfo{person}{Robert~Adams Edward}, \bibinfo{person}{Mike Dvorkin},
  \bibinfo{person}{Youcef Laribi}, \bibinfo{person}{Vijoy Pandey},
  \bibinfo{person}{Pankaj Garg}, {and} \bibinfo{person}{Nik Weidenbacher}.}
  \bibinfo{year}{2016}\natexlab{}.
\newblock \bibinfo{booktitle}{\emph{{OpFlex Control Protocol}}}.
\newblock \bibinfo{type}{Internet Draft} draft-smith-opflex-03.
  \bibinfo{institution}{Internet Engineering Task Force}.
\newblock


\bibitem[Stoenescu et~al\mbox{.}(2018)]%
        {StoenescuDPNR18}
\bibfield{author}{\bibinfo{person}{Radu Stoenescu}, \bibinfo{person}{Dragos
  Dumitrescu}, \bibinfo{person}{Matei Popovici}, \bibinfo{person}{Lorina
  Negreanu}, {and} \bibinfo{person}{Costin Raiciu}.}
  \bibinfo{year}{2018}\natexlab{}.
\newblock \showarticletitle{Debugging {P4} programs with vera}. In
  \bibinfo{booktitle}{\emph{Proceedings of the 2018 Conference of the {ACM}
  Special Interest Group on Data Communication}}. \bibinfo{pages}{518--532}.
\newblock


\bibitem[Talbi(2009)]%
        {Talbi2009}
\bibfield{author}{\bibinfo{person}{El-Ghazali Talbi}.}
  \bibinfo{year}{2009}\natexlab{}.
\newblock \bibinfo{booktitle}{\emph{{M}etaheuristics: From design to
  implementation} (\bibinfo{edition}{1} ed.)}.
\newblock \bibinfo{publisher}{John Wiley \& Sons}.
\newblock


\bibitem[Wang et~al\mbox{.}(2017)]%
        {Wang2017}
\bibfield{author}{\bibinfo{person}{Tao Wang}, \bibinfo{person}{Fangming Liu},
  {and} \bibinfo{person}{Hong Xu}.} \bibinfo{year}{2017}\natexlab{}.
\newblock \showarticletitle{An Efficient Online Algorithm for Dynamic {SDN}
  Controller Assignment in Data Center Networks}.
\newblock \bibinfo{journal}{\emph{{IEEE/ACM} Transactions on Networking}}
  \bibinfo{volume}{25} (\bibinfo{year}{2017}), \bibinfo{pages}{2788--2801}.
\newblock


\bibitem[Witten et~al\mbox{.}(2016)]%
        {WittenFH16}
\bibfield{author}{\bibinfo{person}{Ian~H. Witten}, \bibinfo{person}{Eibe
  Frank}, \bibinfo{person}{Mark~A. Hall}, {and} \bibinfo{person}{Christopher~J.
  Pal}.} \bibinfo{year}{2016}\natexlab{}.
\newblock \bibinfo{booktitle}{\emph{Data mining: practical machine learning
  tools and techniques} (\bibinfo{edition}{4} ed.)}.
\newblock \bibinfo{publisher}{Elsevier}.
\newblock


\bibitem[Woo et~al\mbox{.}(2018)]%
        {WooLKS18}
\bibfield{author}{\bibinfo{person}{Seungwon Woo}, \bibinfo{person}{Seungsoo
  Lee}, \bibinfo{person}{Jinwoo Kim}, {and} \bibinfo{person}{Seungwon Shin}.}
  \bibinfo{year}{2018}\natexlab{}.
\newblock \showarticletitle{{RE-CHECKER:} Towards Secure RESTful Service in
  Software-Defined Networking}. In \bibinfo{booktitle}{\emph{Proceedings of the
  2018 {IEEE} Conference on Network Function Virtualization and Software
  Defined Networks}}. \bibinfo{pages}{1--5}.
\newblock


\bibitem[Zalewski(2016)]%
        {Zalewski:AFL}
\bibfield{author}{\bibinfo{person}{Michał Zalewski}.}
  \bibinfo{year}{2016}\natexlab{}.
\newblock \bibinfo{title}{{A}merican {F}uzzy {L}op --- Whitepaper}.
\newblock
\newblock
\urldef\tempurl%
\url{https://lcamtuf.coredump.cx/afl/technical_details.txt}
\showURL{%
\tempurl}


\bibitem[Zhang(2017)]%
        {Zhang17}
\bibfield{author}{\bibinfo{person}{Peng Zhang}.}
  \bibinfo{year}{2017}\natexlab{}.
\newblock \showarticletitle{Towards rule enforcement verification for software
  defined networks}. In \bibinfo{booktitle}{\emph{Proceedings of the 2017
  {IEEE} Conference on Computer Communications}}. \bibinfo{pages}{1--9}.
\newblock


\bibitem[Zhao et~al\mbox{.}(2019)]%
        {ZhaoLWSH19}
\bibfield{author}{\bibinfo{person}{Hui Zhao}, \bibinfo{person}{Zhihui Li},
  \bibinfo{person}{Hansheng Wei}, \bibinfo{person}{Jianqi Shi}, {and}
  \bibinfo{person}{Yanhong Huang}.} \bibinfo{year}{2019}\natexlab{}.
\newblock \showarticletitle{{SeqFuzzer}: An Industrial Protocol Fuzzing
  Framework from a Deep Learning Perspective}. In
  \bibinfo{booktitle}{\emph{Proceedings of the 12th {IEEE} Conference on
  Software Testing, Validation and Verification}}. \bibinfo{pages}{59--67}.
\newblock


\end{thebibliography}

\end{document}